\documentclass[final,5p,times,twocolumn,numbers,sort&compress]{elsarticle}

\usepackage{graphicx}
\usepackage{amsmath}
\usepackage{subfigure}
\usepackage{wrapfig}
\usepackage{epsfig}
\usepackage{float}
\usepackage{color}
\usepackage[section]{placeins}

\newcommand\fverb{\setbox\fverbbox=\hbox\bgroup\verb}
\newcommand\fverbdo{\egroup\medskip\noindent%
			\fbox{\unhbox\fverbbox}\ }
\newcommand\fverbit{\egroup\item[\fbox{\unhbox\fverbbox}]}
\newbox\fverbbox

\newcommand{\be}{\begin{equation}}
\newcommand{\ee}{\end{equation}}
\newcommand{\bea}{\begin{eqnarray}}
\newcommand{\eea}{\end{eqnarray}}
\newcommand{\nn}{\nonumber}

\newcommand{\comment}[1]{}

\def\half{{\textstyle{1\over2}}}

\def\MeV{\mathop{\rm MeV}\nolimits}
\def\GeV{\mathop{\rm GeV}\nolimits}
\def\Tr{{\sf Tr}}
\def\Re{{\sf Re}}
\def\Im{{\sf Im}}
\def\Det{{\sf Det}}
\def\bar{\overline}
\def\hat{\widehat}
\def\tilde{\widetilde}

\def\half{{\scriptstyle \raise.15ex\hbox{${1\over2}$}}}

\newcommand{\beq}{\begin{equation}}
\newcommand{\eeq}{\end{equation}}

\newcommand{\real}{\relax{\rm I\kern-.18em R}}

\def\vek#1{{\bf #1}}
\def\overbar{\overline}
\def\vv{{\bf V}}
\def\rd{{\rm d}}
\def\mco{\multicolumn}

\usepackage{graphicx}

\usepackage{afterpage,float}

\usepackage{amssymb}

\journal{Physics Letters B}

\begin{document}

\begin{frontmatter}

\def\MeV{\mathop{\rm MeV}\nolimits}
\def\GeV{\mathop{\rm GeV}\nolimits}
\def\Tr{{\sf Tr}}
\def\Re{{\sf Re}}
\def\Im{{\sf Im}}
\def\Det{{\sf Det}}
\def\bar{\overline}
\def\hat{\widehat}
\def\tilde{\widetilde}

\def\half{{\scriptstyle \raise.15ex\hbox{${1\over2}$}}}

\newcommand{\Cite}[1]{$\,$\cite{#1}}

\def\vek#1{{\bf #1}}
\def\overbar{\overline}
\def\vv{{\bf V}}
\def\rd{{\rm d}}
\def\mco{\multicolumn}

\title{Can the  nearly conformal sextet gauge model hide the Higgs impostor?}
 
 \author[wupi,budapest]{Zolt\'{a}n Fodor}
 
 \author[uop]{Kieran Holland}
 
 \author[ucsd]{Julius Kuti\corref{cor1}}
 \ead{jkuti@ucsd.edu}
 
 \author[budapest]{D\'{a}niel N\'{o}gr\'{a}di}
 
 \author[livermore]{Chris Schroeder}

 \author[ucsd]{Chik Him Wong}

 \cortext[cor1]{Corresponding author}
 
 \address[wupi]{Department of Physics, University of Wuppertal, 
Gaussstrasse 20, D-42119, Germany\\
J\"ulich Supercomputing Center, Forschungszentrum, 
        J\"ulich, D-52425 J\"ulich, Germany}
 \address[uop]{Department of Physics, University of the Pacific, 
3601 Pacific Ave, Stockton CA 95211, USA\\
Institute for Theoretical Physics, 
Bern University, Sidlerstrasse 5, CH-3012 Bern, Switzerland}
 \address[ucsd]{Department of Physics 0319, University of California, San Diego, 
9500 Gilman Drive, La Jolla, CA 92093, USA}
\address[budapest]{Institute for Theoretical Physics, E\"otv\"os University, 
        H-1117 Budapest, Hungary}
\address[livermore]{Physical Sciences Directorate, Lawrence Livermore National Laboratory,
        Livermore, California 94550, USA}

\vskip -0.2in
\begin{abstract}
New results are reported from large scale lattice simulations of a frequently discussed strongly interacting
gauge theory with a fermion flavor doublet in the two-index symmetric (sextet) representation
of the SU(3) color gauge group. We find that the chiral condensate and the mass spectrum of the sextet model 
are consistent with chiral symmetry breaking in the limit of vanishing fermion mass. 
In contrast, sextet fermion mass deformations of 
spectral properties are not consistent with leading conformal scaling behavior near the critical surface 
of a conformal theory.
A recent paper could not resolve
the conformal fixed point of the gauge coupling from the slowly walking scenario 
of a very small nearly vanishing $\beta$-function~\cite{DeGrand:2012yq}. 
It is argued that overall consistency with our new results is resolved if the sextet model 
is close to the conformal window, staying outside with a very small non-vanishing $\beta$-function. 
The model would exhibit then the simplest composite Higgs mechanism leaving open the possibility
of a light scalar state with quantum numbers of the Higgs impostor. It would emerge as the 
pseudo-Goldstone dilaton state from spontaneous symmetry breaking of scale invariance. 
We will argue that even without association with the dilaton, the scalar Higgs-like state 
can be light very close to the conformal window.
A new Higgs project of sextet lattice simulations is outlined to resolve these important questions.
\end{abstract}

\begin{keyword}
lattice simulations, electroweak sector, technicolor, conformal




\end{keyword}

\end{frontmatter}

\section{Introduction}

The stunning discovery of the 125 GeV Higgs-like particle at the Large Hadron Collider~\cite{:2012gk,:2012gu}
does not exclude new physics beyond the Standard Model (BSM) in the framework 
of some new strongly-interacting gauge
theory with a composite Higgs mechanism, an idea which was outside experimental 
reach when it was first introduced as an attractive BSM scenario
~\cite{Weinberg:1979bn,Susskind:1978ms,Dimopoulos:1979es,
Eichten:1979ah,Farhi:1980xs,Holdom:1984sk,Appelquist:1987fc,Miransky:1996pd}. 
The original framework has been considerably  extended by new
explorations of the multi-dimensional theory space in fermion flavor number, the choice of color gauge group, and
fermion representation~\cite{Caswell:1974gg,Banks:1981nn,Marciano:1980zf,Kogut:1984sb,Appelquist:2003hn, Sannino:2004qp,
Dietrich:2005jn,Luty:2004ye,Dietrich:2006cm,Kurachi:2006ej}.
Systematic and non-perturbative lattice studies play an important role in studies of this extended theory 
space~\cite{Fodor:2009wk,Fodor:2011tw,Fodor:2011tu,Fodor:2012uu,Appelquist:2007hu,Appelquist:2009ty,
Appelquist:2011dp,Appelquist:2009ka,Deuzeman:2008sc,
Deuzeman:2009mh,Deuzeman:2011pa,Hasenfratz:2009ea,Hasenfratz:2010fi,Cheng:2011ic,Jin:2009mc,Jin:2010vm,
Aoki:2012kr,Catterall:2007yx,Catterall:2008qk,Hietanen:2008mr,
Hietanen:2009az,DelDebbio:2010hx,Bursa:2010xn,DelDebbio:2010ze,DelDebbio:2011kp, 
Shamir:2008pb,DeGrand:2010na,DeGrand:2011cu,Kogut:2010cz,Kogut:2011ty, Bilgici:2009kh,
Itou:2010we,Yamada:2009nt,Hayakawa:2010yn,Gavai:1985wi,Attig:1987mf,
Meyer:1990xd,Damgaard:1997ut,Kim:1992pk,Brown:1992fz,Iwasaki:2003de}.  
Even without spin and parity information, the new Higgs-like particle with decay modes 
not far from that of
the Standard Model brings new focus and clarity to the search for the proper theoretical
framework.
%

One example is the light dilaton as a pseudo-Goldstone particle of spontaneous  breaking of 
scale invariance that has been featured in recent phenomenological discussions as a viable
interpretation of the discovery~\cite{Ellis:2012hz,Low:2012rj}. 
Nearly conformal 
gauge theories serve as  theoretical laboratories for realistic implementations
of this scenario ~\cite{Yamawaki:1985zg,Bardeen:1985sm,Holdom:1986ub,Goldberger:2007zk,
Appelquist:2010gy,Grinstein:2011dq,Antipin:2011aa,Hashimoto:2010nw,Matsuzaki:2012fq}. Unfortunately,
a  credible  realization of the idea as a strongly interacting BSM gauge theory 
is still lacking. We investigate here
a candidate theory with a fermion flavor doublet in the two-index symmetric (sextet) representation
of the SU(3) color gauge group close to the conformal window, if it can hide a light  Higgs-like scalar state with or without
dilaton-like interpretation.

The sextet force
and a new fermion doublet driving electroweak symmetry breaking was introduced in QCD a long time ago by Marciano~\cite{Marciano:1980zf}.
Early pioneering lattice work, limited to the quenched approximation at that time,  
investigated the sextet fermion representation~\cite{Kogut:1984sb}.
The main difference in the model we investigate here is the introduction of a new 
SU(3) gauge force not associated with QCD gluons and motivated by ideas of compositeness from a new super-strong force.
After chiral symmetry breaking we find three massless Goldstone pions in the spectrum
providing the minimal realization of the Higgs
mechanism, just like in the original technicolor idea~\cite{Weinberg:1979bn,Susskind:1978ms}.
The important new ingredient is the sextet representation of the fermion doublet which brings the model very 
close to the conformal window as indicated in a recent paper~\cite{DeGrand:2012yq}. 
The accuracy of the very small nearly vanishing $\beta$-function in difficult simulations could not resolve
the existence of a conformal fixed point gauge coupling from the alternative slowly walking scenario.
When combined with our 
observation of  chiral symmetry breaking ($\chi{\rm SB}$) reported here for small fermion mass deformations,
the overall consistency of all simulations is resolved if the sextet model 
is close to the conformal window with a very small non-vanishing $\beta$-function (see, also~\cite{Kogut:2010cz,Kogut:2011ty}). 
In this case the model exhibits the simplest composite Higgs mechanism and leaves open the possibility
of a light scalar state with quantum numbers of the Higgs impostor emerging as the 
pseudo-Goldstone dilaton state from spontaneous symmetry breaking of scale invariance. 
Even if scale symmetry breaking is entangled with
$\chi{\rm SB}$  without dilaton interpretation, 
a light Higgs-like scalar state can emerge  from the new force close to the conformal window.
Our new Higgs project with  lattice simulations in the sextet model may resolve these important problems.

In section 2 we will outline the computational strategy and the simulation set-up including the important treatment
of finite size effects. In section 3 results on the chiral condensate are presented with extrapolation to the massless
fermion limit. The spectrum is presented in Section 4 and compared with the $\chi{\rm SB}$ hypothesis.
In Section 5 it is shown that fermion mass deformations of 
spectral properties are not consistent in the model with conformal scaling behavior near the critical surface 
of a conformal theory. Section 6 will describe the new Higgs project to determine the scalar $0^{++}$ mass spectrum 
when disconnected diagrams are included in the calculations. Closely related to the dilaton interpretation, we also outline 
in Section 6 the important role of the non-perturbative gluon
condensate and our strategy for investigating it within our new Higgs project.

\section{Computational strategy and lattice simulations}

Probing $\chi{\rm SB}$, and conformal behavior for comparison, we extrapolate the spectrum 
to infinite volume at fixed fermion mass $m$. In large volumes the leading 
finite size corrections  are exponentially small and dominated by the lowest state of the spectrum which has pion quantum numbers.
From the mass spectrum, extrapolated to  infinite volume, we can probe the pattern of $\chi{\rm SB}$
when small fermion mass deformations are simulated close to the massless limit. We  also probe the hypothesis of
mass deformed conformal scaling behavior. Our results, as we report here, strongly favor the $\chi{\rm SB}$ hypothesis.

\subsection{The algorithm and simulation results}
We have used the tree-level Symanzik-improved gauge action for all simulations reported in this paper.
The conventional $\beta=6/g^2$ lattice gauge coupling is defined as the overall
factor in front of the well-known terms of the Symanzik lattice action.  Its values are $\beta=3.20$ and
$\beta=3.25$ for our simulations.
The link variables in the staggered fermion matrix were exponentially smeared with  two
stout steps~\cite{Morningstar:2003gk}; the precise definition of the staggered stout action was given in~\cite{Aoki:2005vt}.  
The RHMC algorithm was deployed in all runs. The fermion flavor doublet requires rooting in the algorithm.
For molecular dynamics time evolution we applied multiple time scales~\cite{Urbach:2005ji} and the
Omelyan integrator~\cite{Takaishi:2005tz}.
Our error analysis of  hadron masses used correlated fitting with double jackknife procedure on the covariance matrices~\cite{DelDebbio:2007pz}.
The time histories of the fermion condensate, the plaquette, 
and correlators were used to monitor autocorrelation times in the simulations.

We have new simulation results at $\beta=3.2$ in the fermion mass range ${\rm m=0.003-0.010}$ on
$24^3\times48$,  $28^3\times56$,  and $32^3\times64$ lattices. Five fermion masses
at  ${\rm m=0.003,0.004,0.005,0.006,0.008}$ are used in most fits.
A very large and expensive $48^3\times96$ 
run was added recently at ${\rm m=0.003}$ to control finite size effects.
%
%
We also have new simulation results at $\beta=3.25$ in the mass range ${\rm m=0.004-0.008}$ on
$24^3\times48$,  $28^3\times56$, and $32^3\times64$ lattices. 

\subsection{Finite size effects}
Infinite-volume extrapolations of the lowest state in the spectrum 
\begin{figure}[h!]
\begin{center}
\begin{tabular}{c}
\includegraphics[height=5cm]{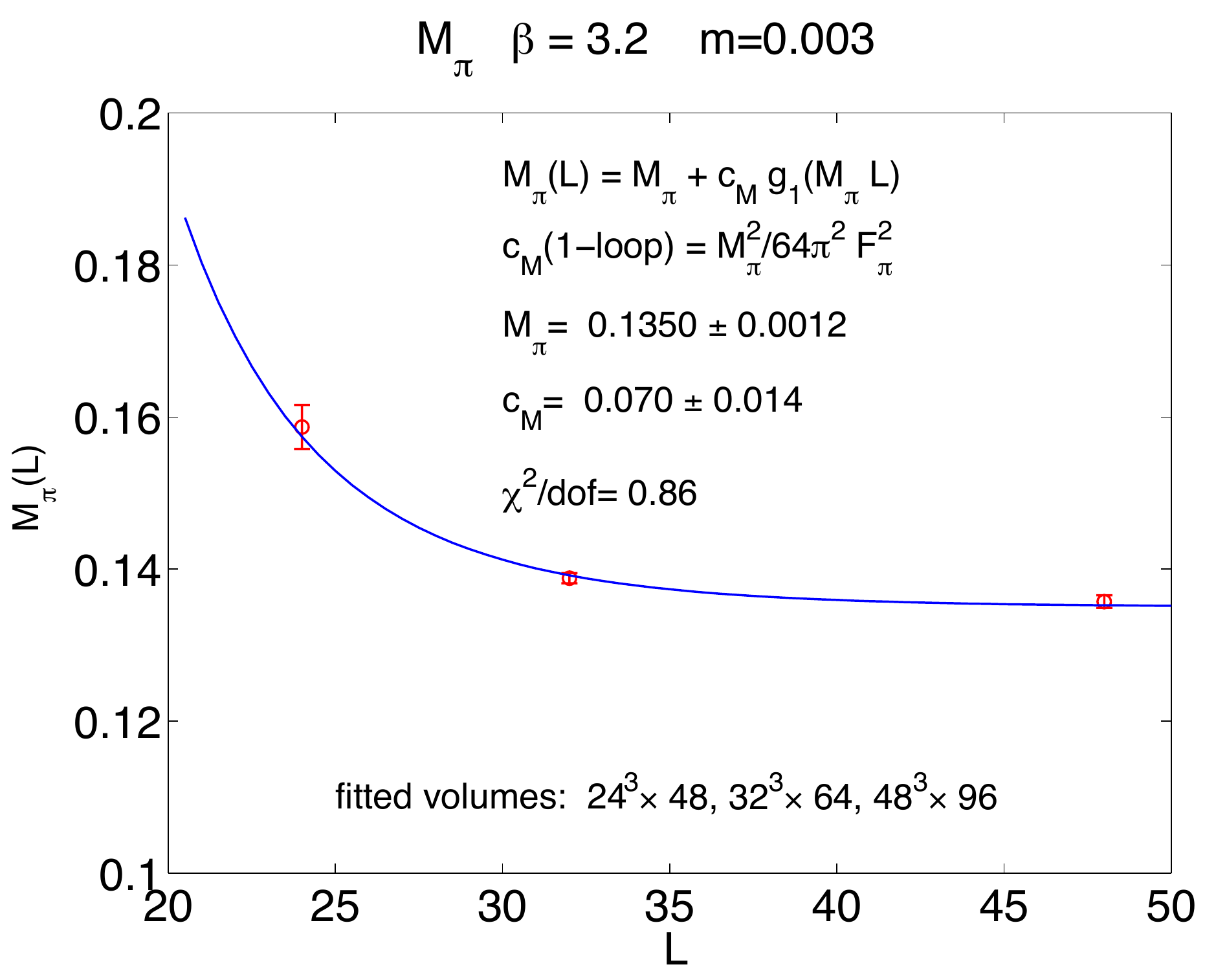}\\
\includegraphics[height=5cm]{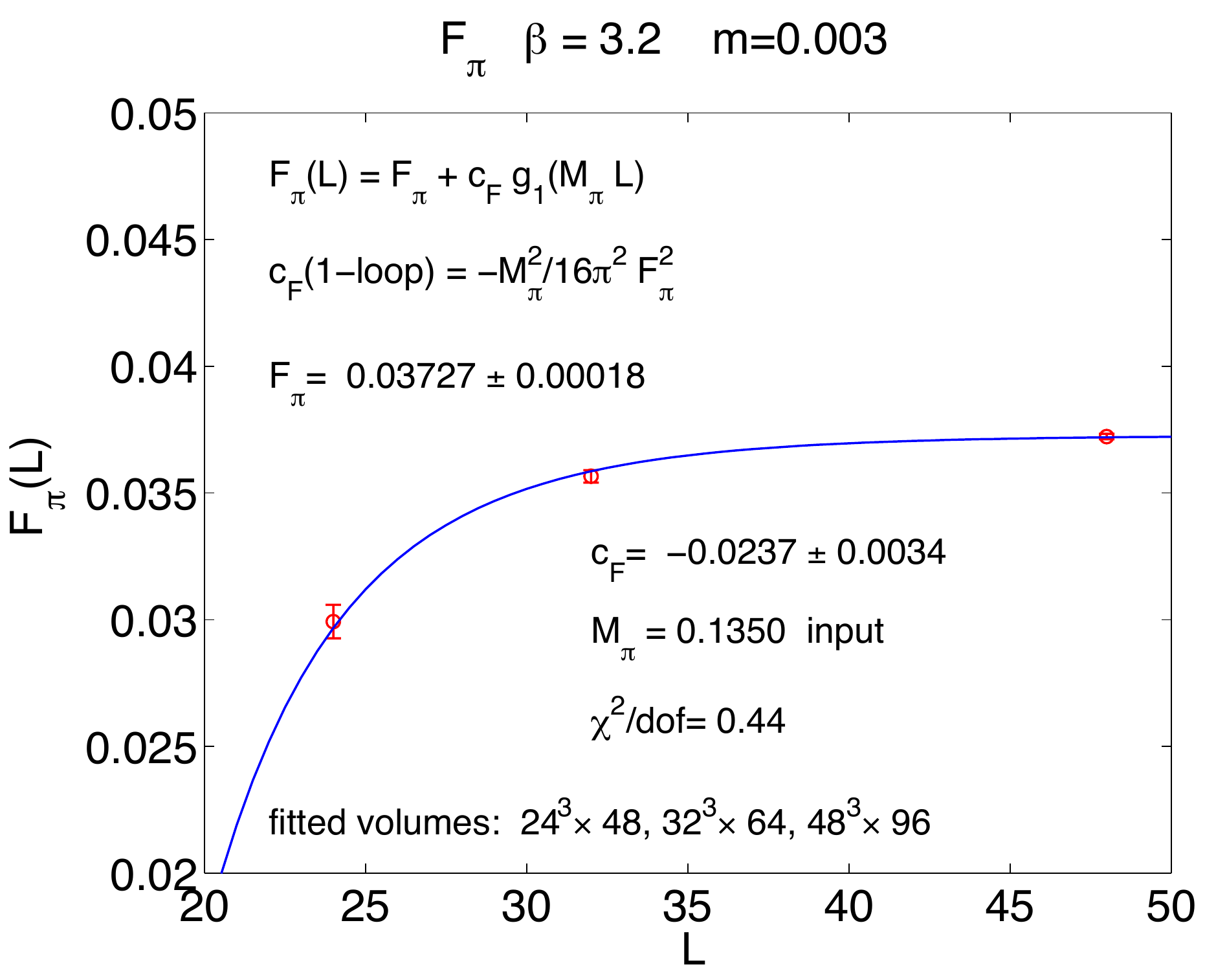}\\
\includegraphics[height=5cm]{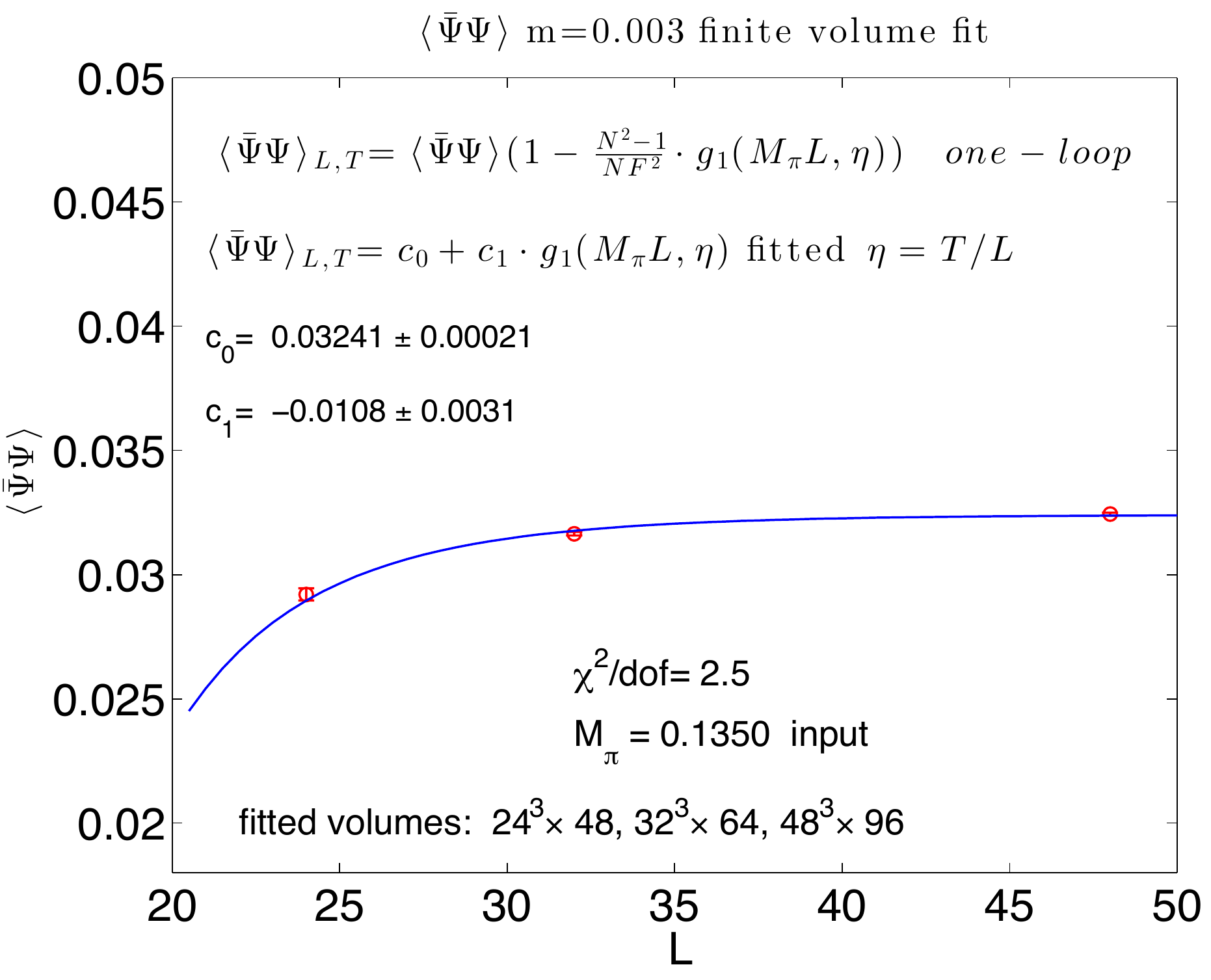}
\end{tabular}
\end{center}
\vskip -0.2in
\caption{\footnotesize  Finite volume dependence at the lowest fermion mass for $\beta=3.2$.
The form of $\tilde g_1(\lambda,\eta)$ is a complicated infinite sum which contains Bessel
functions and requires numerical evaluation~\cite{Gasser:1986vb}. Since we are not in the chiral log regime, the prefactor of
the $\tilde g_1(\lambda,\eta)$ function was replaced by a fitted coefficient. The leading term of  the function
$\tilde g_1(\lambda,\eta)$ is a special exponential Bessel function $K_1(\lambda)$ which dominates in the simulation range.}
\vskip -0.1in
\label{fig:sextetInfVol}
\end{figure}
with pion quantum numbers, the related $F_\pi$,
and the condensate $\langle\bar\psi\psi\rangle$
are shown in Figure~\ref{fig:sextetInfVol} where $\tilde g_1(\lambda,\eta)$ describes finite volume corrections 
from the exchange of the lightest pion state with  $\lambda=M_\pi L$ and lattice aspect ratio $\eta=T/L$,
similarly to what was introduced in~\cite{Leutwyler:1987ak}.  
The fitting procedure approximates the leading treatment of  the pion which wraps around the finite volume,
whether in chiral perturbation theory ($\chi{\rm pt}$), or in L\"uscher's non-perturbative finite size analysis~\cite{Luscher:1985dn}. 
This equivalence relaxes the requirement on the fitted parameters $c_M$,$c_F$,$c_1$ to agree with 1-loop  
$\chi{\rm PT}$ as long as the pion is the lightest state dominating the finite volume corrections.
It should be noted that the form of the fitting function $\tilde g_1(\lambda,\eta)$ does not commit to the chirally broken phase.
At fixed fermion mass $m$, the leading exponential term of the function is also the expected  behavior 
in the conformal phase with mass deformation. The asymptotic exponential form simply originates from the 
lightest state wrapping around 
the volume once emitted from and re-absorbed by the composite state whose sensitivity to finite volume corrections is 
being investigated. The analysis is therefore applicable to both mass deformed phases with different symmetry properties.

The infinite-volume limits of $M_\pi$, $F_\pi$,  and $\langle\bar\psi\psi\rangle$ for  $m=0.003$ at $\beta=3.2$
were determined self-consistently from the fitting procedure. Similar fits were applied to other composite states.
The value of $M_\pi$ in the fit of the top plot in Figure~\ref{fig:sextetInfVol} was 
determined from the highly non-linear fitting function and used as input in the other two fits.
Based on the fits at $m=0.003$,  
the results are within one percent of the infinite-volume  limit at $M_\pi L= 5$.
In the fermion mass range $m \geq 0.004$ the condition $M_\pi L> 5$ is reached at $L=32$.
Although it will require high precision runs to test, we expect less than one percent residual finite size effects
in the  $32^3\times64$ runs for $m \geq 0.004$.
Based on these observations, we will interpret the results from the  $32^3\times64$ runs for $m \geq 0.004$
as infinite-volume  behavior in mass deformed chiral and conformal analysis.

\section{The chiral condensate} 

Our simulations show that the chiral condensate $\langle \bar{\psi}\psi\rangle$ is consistent with $\chi{\rm SB}$ 
and remains non-vanishing in the massless fermion limit.
It has the infinite-volume spectral representation,
\be
\langle \bar{\psi}\psi\rangle = -2m\cdot\int^{\Lambda}_0 d\lambda\frac{\rho(\lambda)}{m^2+\lambda^2}\, ,
\ee
which is UV-divergent when the cutoff $\Lambda$ is taken to  infinity. 
The divergences are isolated by writing the integral of the spectral representation in twice subtracted form~\cite{Leutwyler:1992yt},  
\bea
&&\langle \bar{\psi}\psi\rangle = -2m\cdot\int^{\mu}_0 d\lambda\frac{\rho(\lambda)}{m^2+\lambda^2}\nn\\
&&~~~~~~~~~~~ -2m^5\cdot\int^{\Lambda}_\mu\frac{d\lambda}{\lambda^4}\frac{\rho(\lambda)}{m^2+\lambda^2}
+c_1(a)\cdot m + c_3(a)\cdot m^3 \, .
\label{eq:condensate}
\eea
The first integral in Eq.~(\ref{eq:condensate}) isolates the infrared part and recovers the well-known relation 
$\langle \bar{\psi}\psi\rangle=-\pi\rho(0) $ in the $m\rightarrow 0$ limit~\cite{Banks:1979yr}. 
The linear fermion mass term $c_1(a)\cdot m$ is a quadratically divergent UV contribution 
$\approx a^{-2}\cdot m$ with lattice cutoff $a$.
There is also a very small
third-order UV term  $c_3(a)\cdot m^3$ without power divergences which is hard to detect for small $m$ and has 
not been tested within the accuracy of the simulations. 

IR finite contributions to the condensate from the chiral Lagrangian are connected at the low energy scale $\mu$ with the
first integral in Eq.~(\ref{eq:condensate}). In the chiral expansion of the condensate there is an $m$-independent constant
term which is proportional to $ B F^2$, a linear term proportional to $ B^2\cdot m$, a quadratic term $\sim B^3F^{-2}\cdot m^2$, and higher 
order terms, in addition to
logarithmic corrections generated from chiral loops. 
The expansion in the fermion mass is expressed in terms of low energy constants 
of chiral perturbation theory, like $B$ and $F$~\cite{Bijnens:2009qm}.

\begin{figure}[h!]
\begin{center}
\begin{tabular}{c}
\includegraphics[width=7cm]{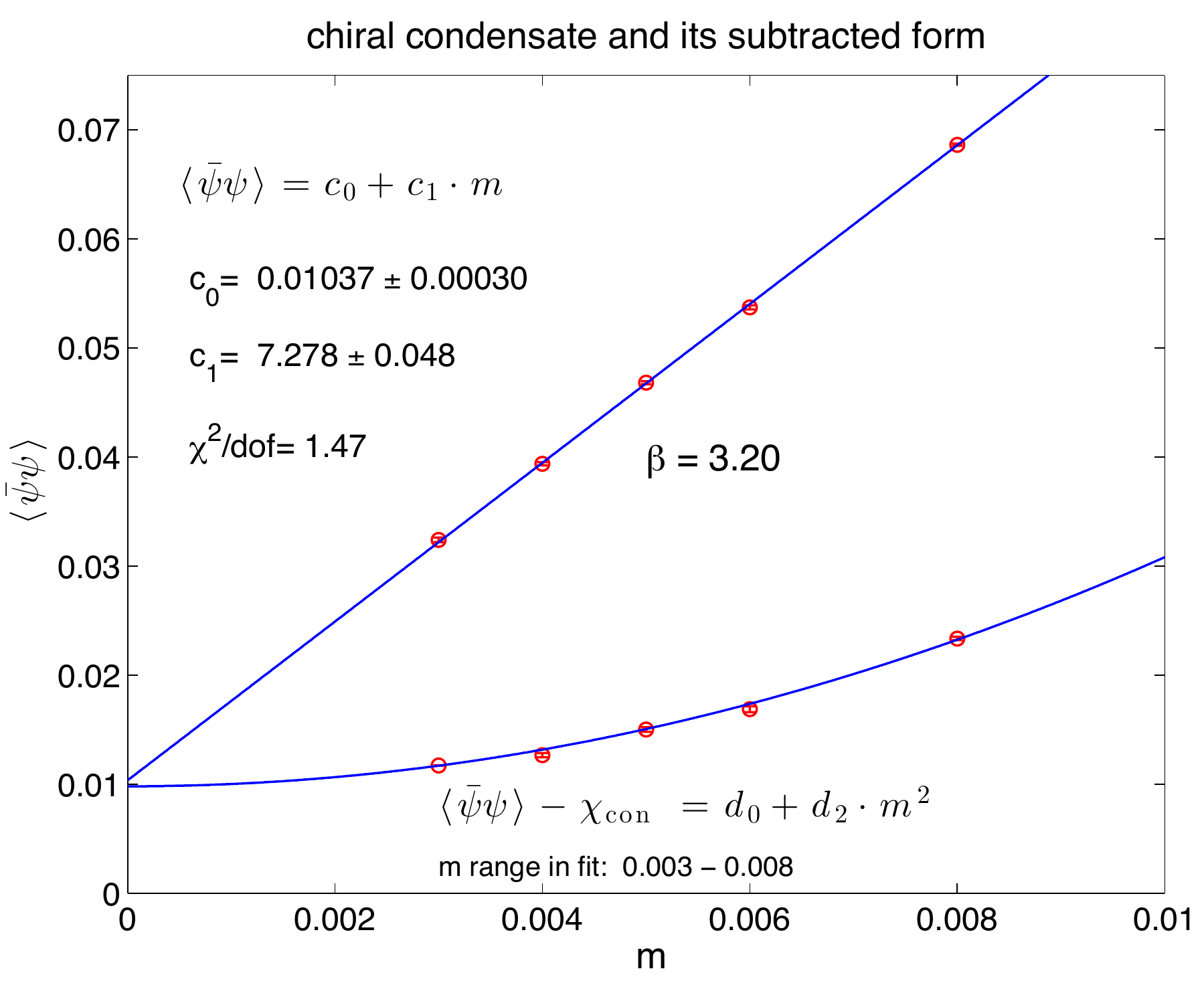}\\
\includegraphics[width=7.2cm]{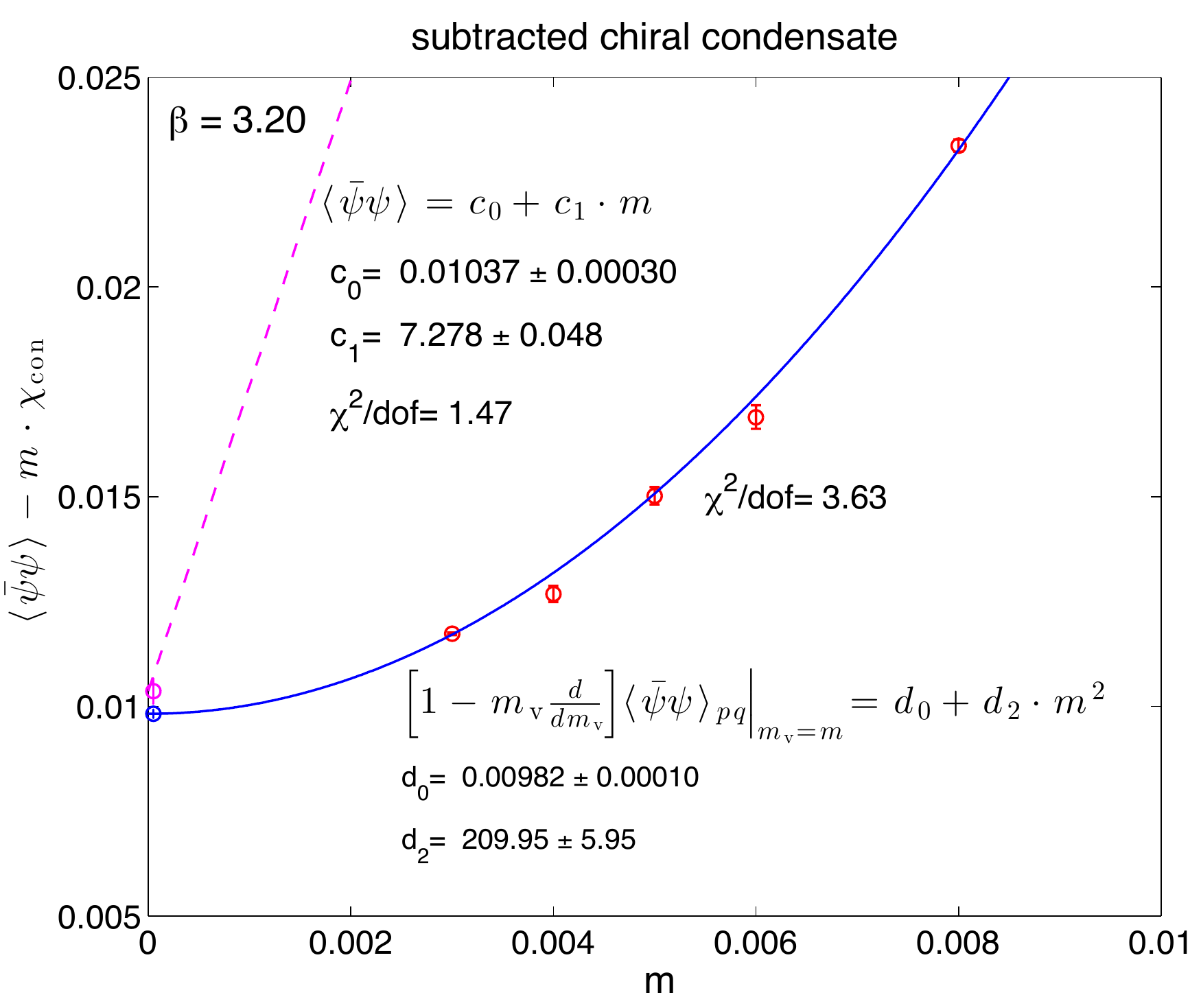}
\end{tabular}
\end{center}
\vskip -0.2in
\caption{\footnotesize The chiral condensate and its reduced form with subtracted derivative  (both have to converge  
to the same chiral limit) are shown in the top plot with linear fit to the condensate. The data without derivative subtraction cannot
detect higher order fermion mass terms with significant accuracy. 
 The fit to the reduced form
with subtracted derivative is defined in the text and shown in the magnified lower plot. 
A linear term is not included in this fit since the subtracted derivative form approximately eliminates it. 
The value of  $d_0$ at $m=0$  is shown to be consistent
with the direct determination of $c_0$ from the chiral limit of $\langle \bar{\psi}\psi\rangle$.
The consistency is very reassuring since the two results are derived from independent determinations.
For $m=0.003$ the data from infinite-volume extrapolation were used in the fit. 
As we explained earlier, at higher $m$ values the largest volume 
$32^3\times 64$ runs were used for the condensate and its derivative subtraction.}
\label{fig:PbPsextet}
\end{figure}
We used two independent methods for the determination of the chiral condensate in the
massless fermion limit. In the first method fits were made directly to 
 $\langle \bar{\psi}\psi\rangle$ with constant and linear terms in the fitted function.
Quadratic 
and third order  terms are hard to detect within the accuracy of the data. 
The result is shown in the top plot of Figure~\ref{fig:PbPsextet}.  
When the quadratic term is added to the fit,
the massless intercept  $c_0=\langle \bar{\psi}\psi\rangle_{m=0}$ from the quadratic fit agrees with the one from the linear fit 
and the quadratic fit coefficient in $c_2\cdot m^2$
is zero within fitting error. 

For an independent determination, we also studied the subtracted chiral condensate operator defined with the help
of the connected part $\chi_{conn}$ of the chiral susceptibility $\chi$,
\bea
&&\Bigl [1-m_{\rm v}\frac{ d}{dm_{\rm v}}\Bigr ] \langle\bar\psi\psi\rangle\Big |_{m_{\rm v}=m}
 = \langle\bar\psi\psi\rangle - m\cdot\chi_{con}~,\nn \\
&& \chi =\frac{ d}{dm} \langle\bar\psi\psi\rangle = \chi_{con} + \chi_{disc}~, 
~~\chi_{con}=\frac{ d}{dm_{\rm v}}\langle\bar\psi\psi\rangle_{pq}\Big |_{m_{\rm v}=m} .
\eea
The derivatives  $d/dm$ and  $d/dm_{\rm v}$ are taken at fixed gauge coupling $\beta$. The derivative
$d/dm_{\rm v}$ is defined in the partially quenched functional integral of $\langle \bar{\psi}\psi\rangle_{pq}$
with respect to the valence mass $m_{\rm v}$
and the limit $m_{\rm v}=m$ is taken after differentiation.
The removal of the derivative term significantly reduces the 
dominant linear part of the $\langle \bar{\psi}\psi\rangle$ condensate without changing the intercept in the $m=0$ limit. 
Once the derivative term is subtracted, the first non-perturbative IR contribution, 
quadratic in $m$, is better exposed. 
The two independent determinations give consistent non-vanishing fit results in the massless chiral limit
as shown in  the lower plot of Figure~\ref{fig:PbPsextet}.

The independent determinations of the non-vanishing condensate  in the chiral limit with separate fits $c_0=\langle \bar{\psi}\psi\rangle_{m=0}$ 
and $d_0=\langle \bar{\psi}\psi\rangle_{m=0}$ are  consistent with each other but differ
from the GMOR~\cite{GellMann:1968rz} relation  $\langle\bar{\psi}\psi\rangle=2BF^2$ by a factor of two. 
As shown in the next section, the value of $2B$ is determined in lattice units from the pion spectrum using the leading
$M^2_\pi=2B\cdot m$ relation. We find the numerical value 
$2B=  6.35(21)$ as shown in the top plot of Figure~\ref{fig:MpiFpi}. 
$F$ is determined from the pseudoscalar correlator which satisfies the PCAC relation. We find in lattice units 
the numerical value $F=0.0279(4)$ from the lower plot of Figure~\ref{fig:MpiFpi} with $2BF^2 = 0.0049(2)$. 
Both sides of the GMOR relation are sensitive to cutoff effects in $B$ and $F$ at bare lattice coupling $\beta=3.2$.
Our preliminary fits based on staggered chiral perturbation theory 
indicate that cutoff effects modifying the continuum values of $B$ and $F$ are likely sources of the discrepancy~\cite{jk}. 
Some increase in  the cutoff dependent values of $B$ and $F$, which is the observed trend, would bring the two sides of
the GMOR relation in agreement.

\section{Spectral tests of the $\chi {\rm SB}$ hypotheses}

\subsection{Strategy and challenges of the spectrum analysis}

Spectrum calculations in a gauge theory with massless fermions require important and difficult lattice extrapolations:
\begin{enumerate}
\item[(1)]  Extrapolation  from finite lattice size to infinite volume,
\item[(2)]  Extrapolation to the massless fermion limit,
\item[(3)]  Extrapolation in lattice spacing to the continuum.
\end{enumerate}
All three issues will be addressed as we present details of the spectrum analysis in this section. The strategy of 
finite size corrections was explained in Section 2 and it will be applied here. Extrapolation from
finite fermion masses will be used to test the two contrasting hypotheses, one with $\chi{\rm SB}$ 
and the other with conformal behavior. As a first step to address the removal of finite lattice spacing, we will
compare the Goldstone and non-Goldstone pion spectra at two different lattice spacings
to probe the restoration of taste symmetry for staggered fermions as the lattice spacing is decreased.

\subsection{The Goldstone pion and $F_\pi$}

The chiral Lagrangian describes the low energy theory of
Goldstone pions and non-Goldstone pions in the staggered lattice fermion formulation.
It will be used as an effective tool probing the $\chi{\rm SB}$  hypothesis at finite fermion masses
including extrapolation to the massless chiral limit.
\begin{figure}[h!]
\begin{center}
\begin{tabular}{c}
\includegraphics[height=6cm]{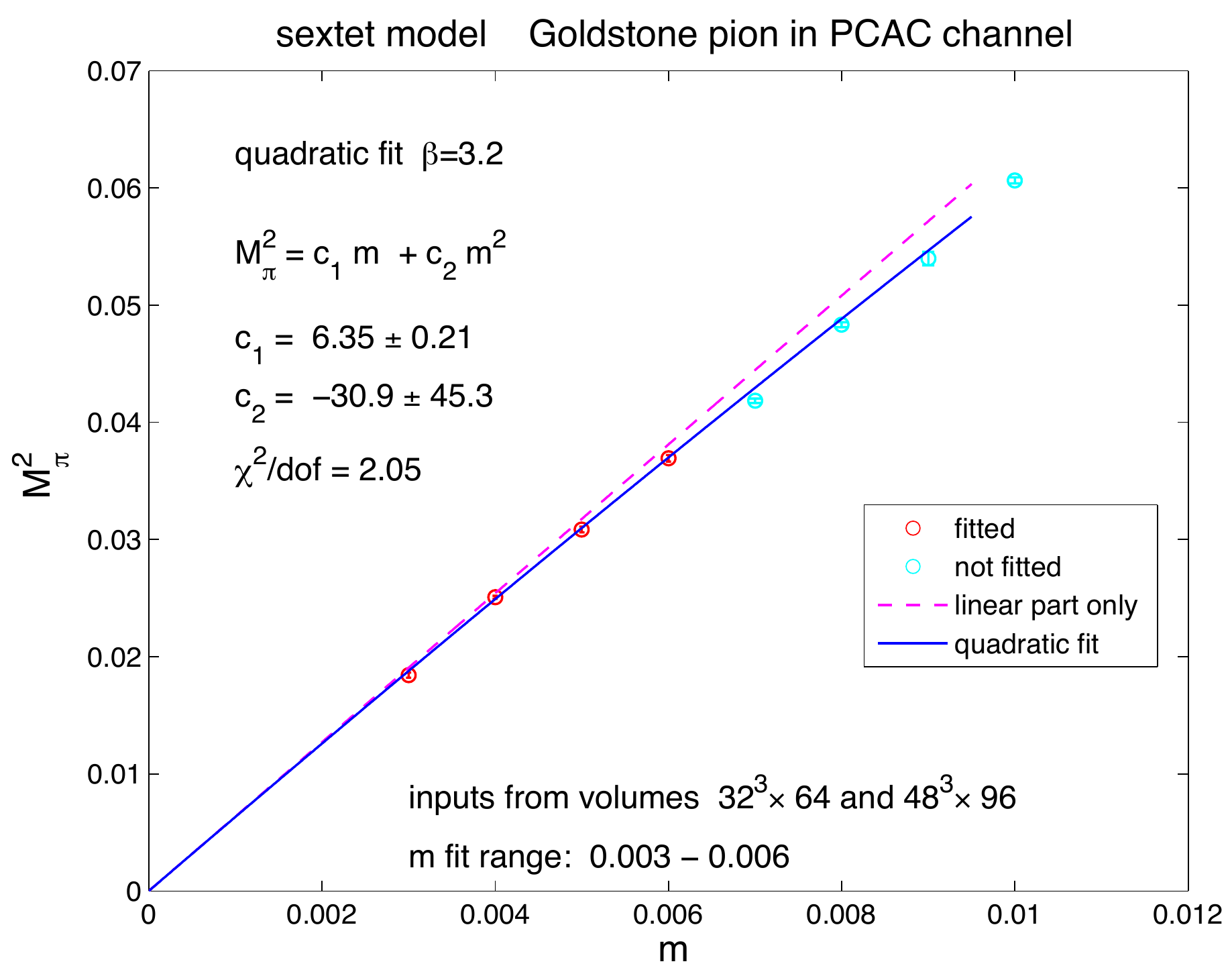}\\
\includegraphics[height=6cm]{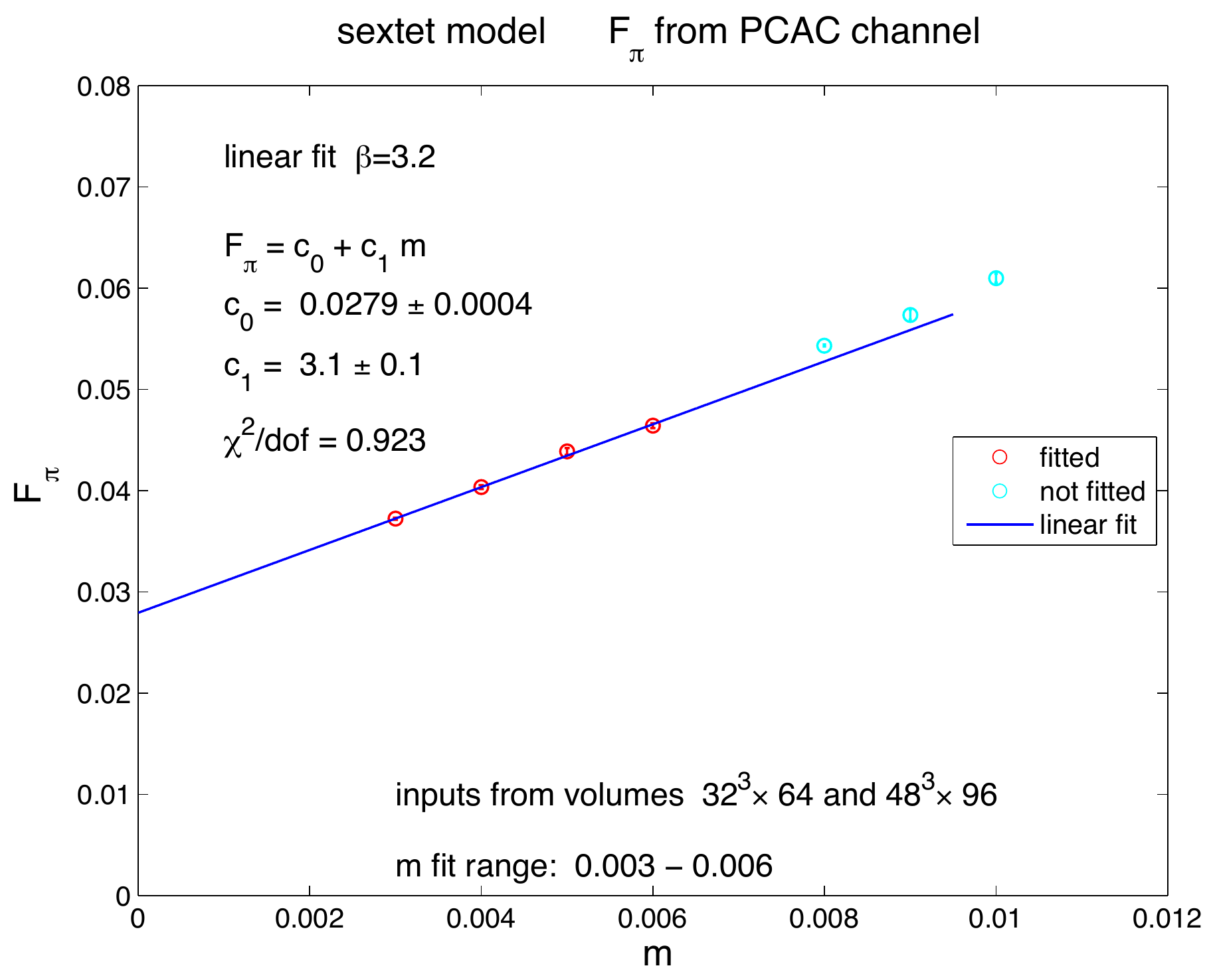}
\end{tabular}
\end{center}
\vskip -0.2in
\caption{\footnotesize  Polynomial fits from the analytic mass dependence of the chiral Lagrangian without logarithmic 
loop corrections are shown for the Goldstone pion and  $F_\pi$. The dashed line in the top plot for the Goldstone pion
shows the leading linear contribution.}
\label{fig:MpiFpi}
\end{figure}
Close to the chiral limit, the  pion spectrum and the pion decay constant $F_\pi$ 
are organized in  powers of the fermion mass $m$ which is an input parameter in the simulations.
Chiral  log corrections to the polynomial terms are
generated from pion loops~\cite{Gasser:1983yg}. Their analysis will
require an extended dateset with high statistics. 

In Section 2 we presented results of infinite-volume extrapolations. The effects are largest 
at $m=0.003$ in our dataset and the infinite-volume limits of $M_\pi$ and  $F_\pi$ were shown for  $m=0.003$ for fixed
lattice cutoff and bare coupling $\beta=3.2$. Similar fits were applied to the chiral condensate and composite states in the
spectrum at $m=0.003$.
Based on the analysis at $m=0.003$,  we determined that the infinite-volume limit is reached at $M_\pi L= 5$ within one percent accuracy.
It is expected that similar or better accuracy is reached for $M_\pi L\geq 5$ at higher $m$ values in all states of the spectrum.
In the fermion mass range $m \geq 0.004$ the condition $M_\pi L> 5$ is reached at $L=32$.
Based on these observations, in fits to the observed pion spectrum and $F_\pi$ 
we will use infinite-volume extrapolation at $m=0.003$ and treat the $32^3\times64$ runs for $m \geq 0.004$
as if the volume were infinite.

In Figure~\ref{fig:MpiFpi} we used the local pion correlator with noisy sources to extract $M_\pi$ and $F_\pi$.
The correlator is tagged as the PCAC channel since the PCAC relation, based on axial Ward identities, holds for this correlator and $F_\pi$ can
be directly determined from the residue of the pion pole.
The other staggered meson states and correlators we use are defined in ~\cite{Ishizuka:1993mt}.
For example, what we call the non-Goldstone scPion and the $f_0$ meson are
identified in correlator I of Table 1 in  ~\cite{Ishizuka:1993mt}. Similarly,  the non-Goldstone i5Pion is from correlator VII, 
the non-Goldstone ijPion is from correlator VIII, and the rho and A1 mesons are
from correlator III of Table 1 in ~\cite{Ishizuka:1993mt}. We measure the Goldstone pion in two different ways, with one of 
them defined above and the other is correlator II of Table 1 in ~\cite{Ishizuka:1993mt}. For baryon states in the sextet fermion  representation,
not presented here,  we
use our own construction of correlators which are different from the baryon correlators of~\cite{Ishizuka:1993mt}. 

Based on the analytic fermion mass dependence of the chiral Lagrangian, and using the lowest four fermion masses,
good polynomial fits were obtained without logarithmic loop corrections as shown in Figure~\ref{fig:MpiFpi} for $M_\pi$ and $F_\pi$. 
Although we could fit $M_\pi$ and $F_\pi$ with the continuum chiral
logarithms included, the two sets of $F$ and $B$ values from separate fits to $M_\pi$ and $F_\pi$ are not quite self-consistent. 
Rooted and partially quenched staggered perturbation theory is a useful procedure 
at finite lattice spacing for simultaneous fits of $M_\pi$ and $F_\pi$ 
with a consistent pair of $F$ and $B$ values~\cite{Aubin:2003mg,Aubin:2003uc}.
The explicit cutoff dependent corrections to the $F$ and $B$ parameters  would require further testing 
at weaker gauge couplings and a set of valence fermion masses. 

We made the first step in this direction by adding a new run set
to our database at $\beta=3.25$. In Figure~\ref{fig:non-GoldstoneSpectrum} we show taste-breaking effects in two pion spectra for comparison.
We find significant reduction in taste breaking at smaller lattice spacing at the weaker coupling. Our
staggered perturbation theory analysis will be presented in a longer follow-up  report
which will also include other results from the new runs at the weaker coupling $\beta=3.25$~\cite{jk}.

\subsection{Taste breaking in the non-Goldstone pion spectrum}

\begin{figure}[h!]
\begin{center}
\begin{tabular}{c}
\includegraphics[height=6cm]{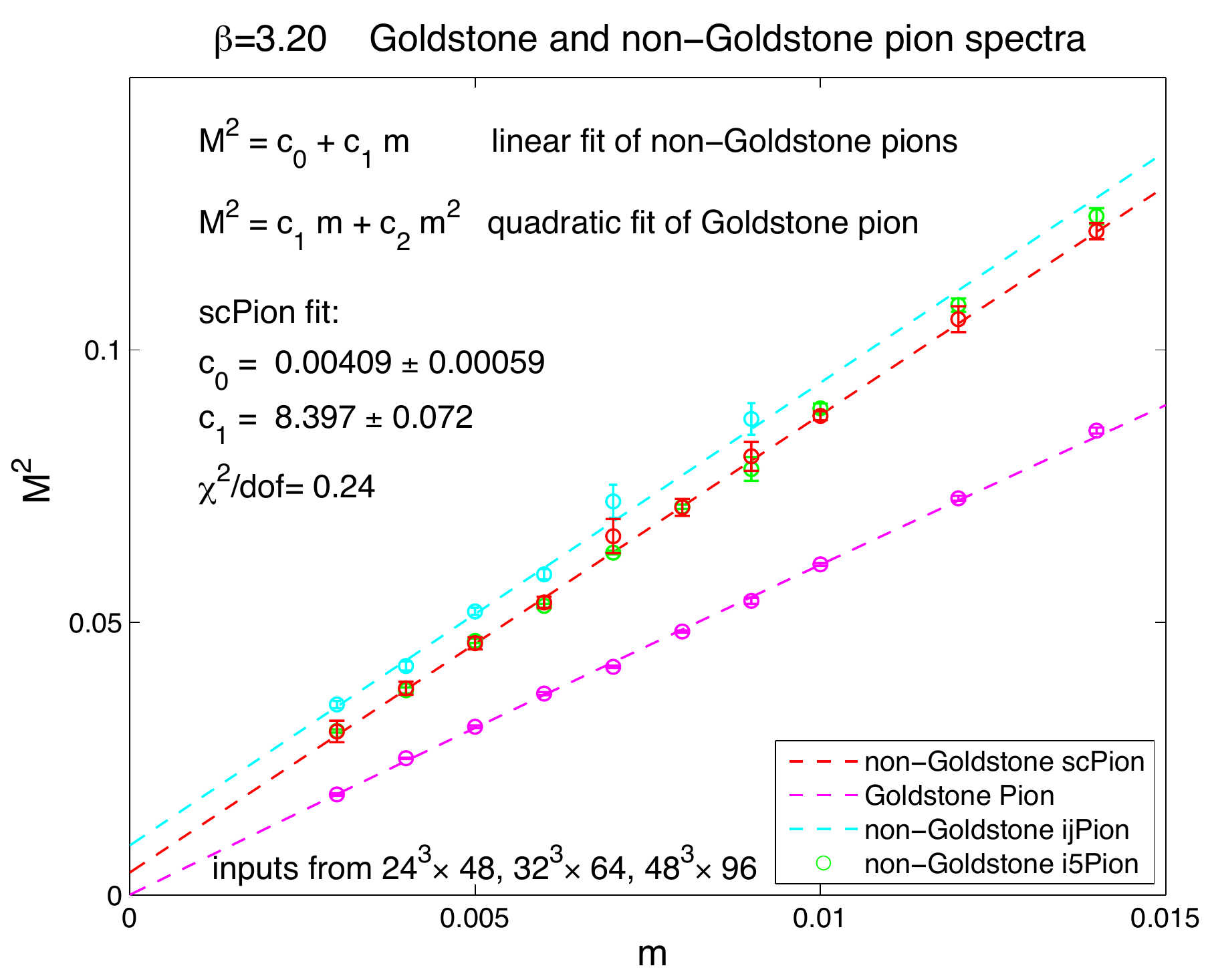}\\
\includegraphics[height=6cm]{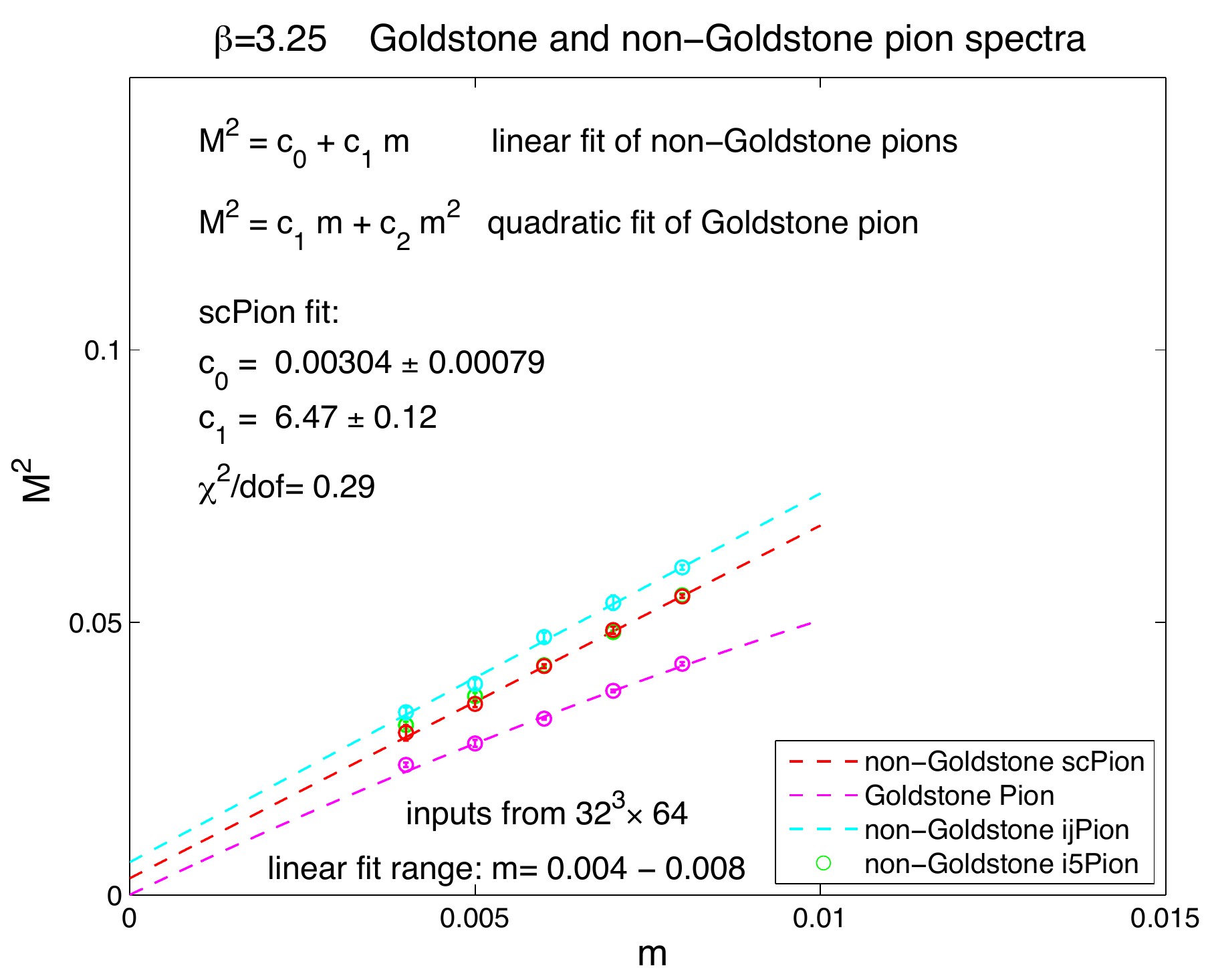}
\end{tabular}
\end{center}
\skip -0.2in
\caption{\footnotesize The top plot in the figure is the spectrum at $\beta=3.2$. It shows the polynomial fit 
of the Goldstone pion  (magenta points). The red points are the non-Goldstone scPion data covering the green i5Pion
data with complete degeneracy. The slightly split ijPion is shown with cyan color.
The lower plot in the figure is the spectrum at $\beta=3.25$. In identical notation it displays the improvement in taste splitting 
with a considerably less taste-broken spectrum when plotted on the same scale.}
\label{fig:non-GoldstoneSpectrum}
\vskip -0.2in
\end{figure}
The non-Goldstone pion spectra, quite different from the one found in QCD, are shown at $\beta=3.2$  in the top plot 
of Figure~\ref{fig:non-GoldstoneSpectrum} using standard notation, introduced earlier. 
The non-Goldstone i5Pion is split from the Goldstone pion and remains exactly degenerate with 
the non-Goldstone scPion, a similar feature in QCD. The new feature is the mass dependence of the split 
between the Goldstone pion and 
the non-Goldstone i5Pion with non-parallel slopes of the fitting functions. 
The non-Goldstone ijPion is further split from the i5Pion with a small mass-independent offset. Although taste breaking 
effects appear substantial on the scale of the plot, they are comparable with those from the HISQ action 
when the lattice spacings are matched~\cite{Bazavov:2010pg}.
The trends of the splits, particularly the fan-out structure and the lack of parallel equi-spaced splits with a constant slope determined by $B$
is characteristic of gauge models as they get close to the conformal window. 
A very small residual mass at $m=0$ is consistent with fits for the non-Goldstone pion states and decreases as 
we lower the lattice spacing with the weaker coupling at $\beta=3.25$.
This is shown in the lower plot of  Figure~\ref{fig:non-GoldstoneSpectrum} which exhibits a similar structure for the same pion states as the top
figure but on a significantly more collapsed scale. Taste breaking is reduced considerably. It will be interesting to conduct a full analysis of 
all data on the finer lattice scale, closer to the continuum limit, and compare with the results presented here on the coarser lattice scale~\cite{jk}.

\subsection{The $ \rho$  and $A_1$ parity partner states}
\vskip -0.003in
It is useful and important to investigate the chiral limit of composite hadron states separated by a gap
from the Goldstone and non-Goldstone pion spectra. The baryon mass gap in the chiral limit can provide further evidence
for $\chi{\rm SB}$ but our preliminary results are not shown here.  
Hadron masses of parity partners also provide important information with split parity masses in the chiral limit.
This is particularly helpful not only to confirm $\chi{\rm SB}$ but to obtain a first estimate on the S parameter 
for probing the model against electroweak precision tests~\cite{Peskin:1991sw}.
As an example, we will briefly review  our results for  the $\rho$ meson state and its parity partner, the ${\rm A_1}$ meson.
Particularly interesting is the 
$\rho-{\rm A_1}$ mass splitting with parity violation.
\begin{figure}[h!]
\begin{center}
\begin{tabular}{c}
\includegraphics[height=5.5cm]{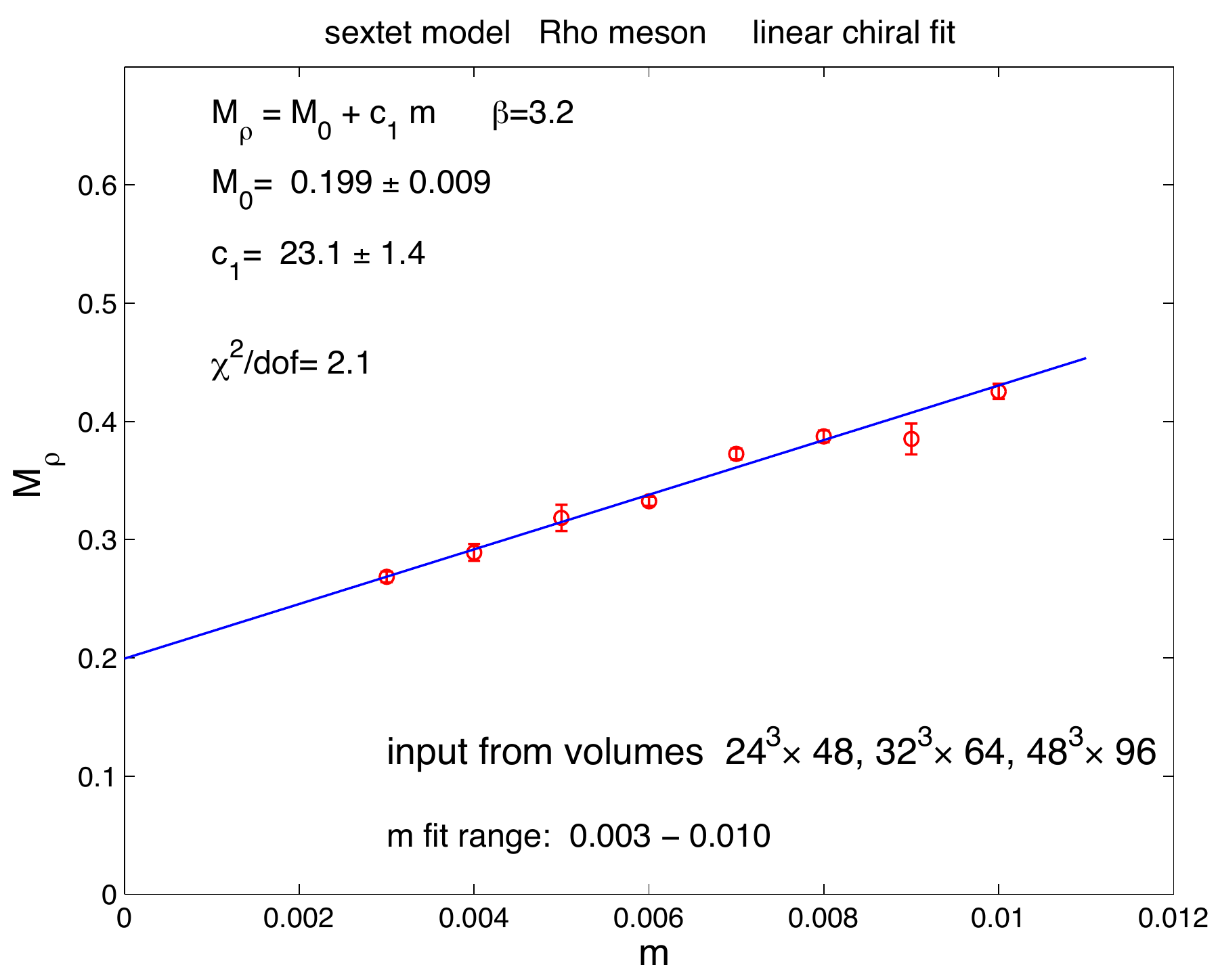}\\
\includegraphics[height=5.5cm]{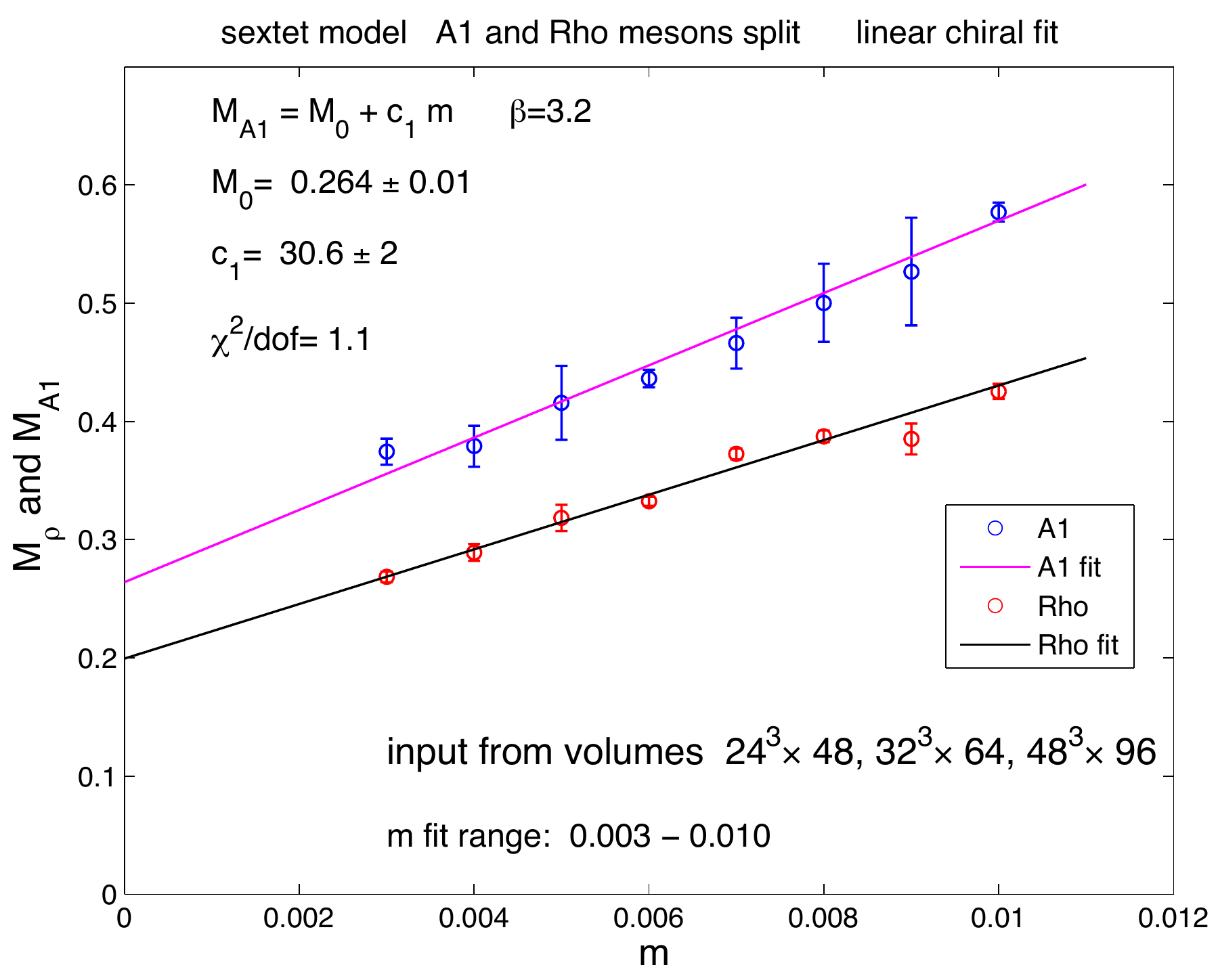}
\end{tabular}
\end{center}
\vskip -0.2in
\caption{\footnotesize  Linear fit to the $\rho$ meson mass is shown in the top plot of the figure. 
The lower plot shows the linear fit to the ${\rm A_1}$ meson superimposed on the $\rho$ meson plot. The parity split
is quite visible with varying size errors in the fitted $m$ range. }
\label{fig:sextetRhoA1}
\vskip -0.1in
\end{figure}
Figure~\ref{fig:sextetRhoA1} shows fits to the $\rho$ meson and its $A_1$ parity partner. The top plot is a linear fit to the
 $\rho$ meson with a non-vanishing mass at $m=0$, consistent with $\chi{\rm SB}$. The lower plot shows the linear fit to the ${\rm A_1}$
meson. Both states extrapolate to non-vanishing masses
in the chiral limit. The split appears to be significant for all fermion masses but the error is too large to resolve the chiral limit. 
More work with higher statistics is needed on this correlator before conclusive
results can be obtained.

\section{Spectral tests of the conformal scaling hypothesis}
Under the conformal scaling hypothesis, the mass $M_\pi$
\begin{figure}[h!]
\begin{center}
\begin{tabular}{c}
\includegraphics[height=5cm]{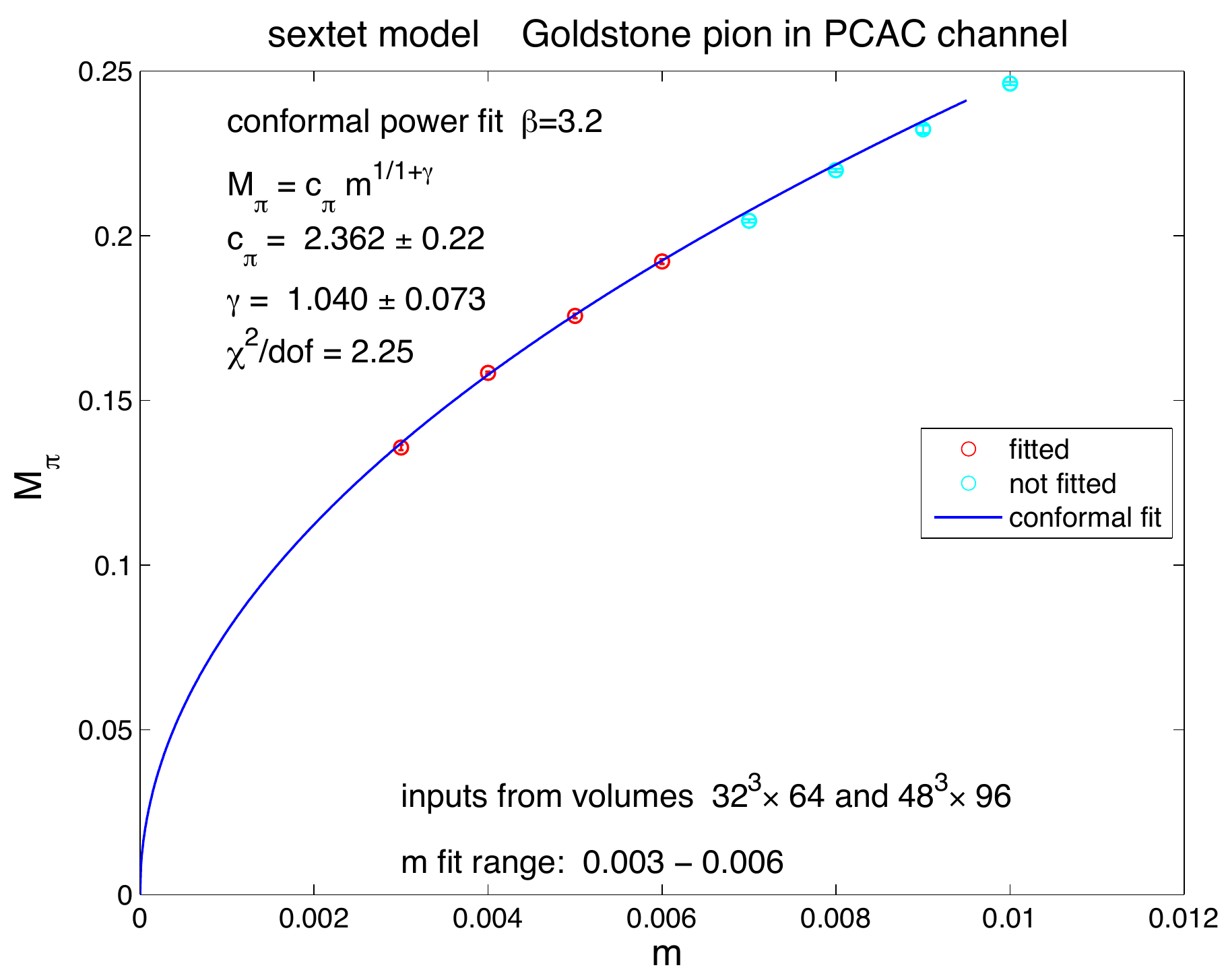}\\
\includegraphics[height=5cm]{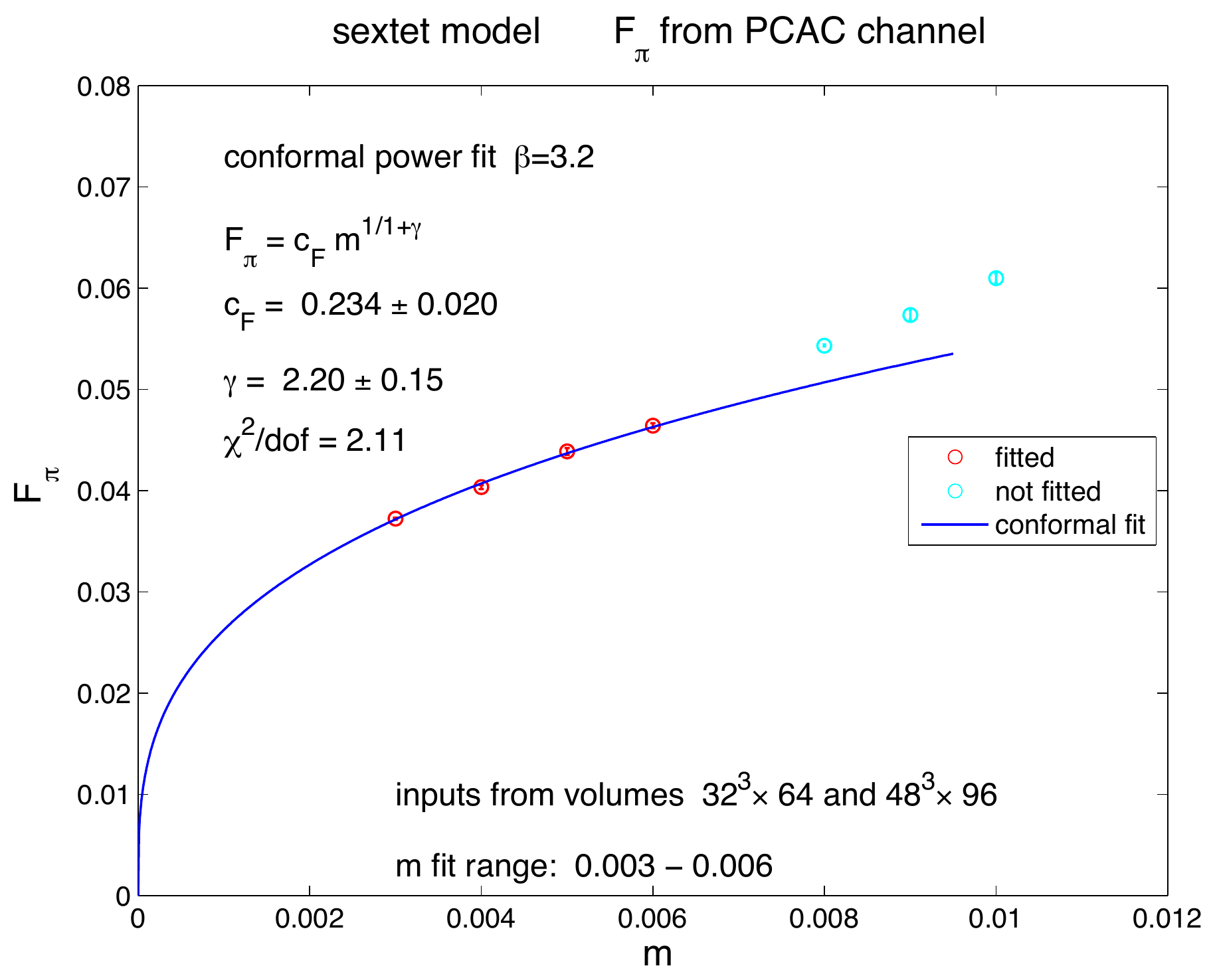}
\end{tabular}
\end{center}
\vskip -0.2in
\caption{\footnotesize The two plots represent separate conformal fits to $M_\pi$ (top) and $F_\pi$ (bottom). 
The separate fits have reasonable $\chi^2$ values but the incompatibility of the fitted $\gamma$ values
disfavors the conformal hypothesis in its leading form.}
%
\label{fig:sextetConformTest1}
\end{figure}
and the decay constant $F_\pi$ are given at leading order by $M_\pi = c_\pi\cdot m^{1/1+\gamma}$ and $F_\pi = c_F\cdot m^{1/1+\gamma}$.  
The coefficients $c_\pi$ and $c_F$
are channel specific but the exponent $\gamma$ is  universal in all channels~\cite{DelDebbio:2010hx,Bursa:2010xn,DelDebbio:2010ze,DelDebbio:2011kp}. 
The leading scaling form sets in for small $m$ values,
close to the critical surface. According to the hypothesis, there is an infrared conformal fixed point on the critical surface which controls
the conformal scaling properties of small mass deformations. 
All masses of the spectrum can be subjected to similar conformal
scaling tests, but we will mostly focus on accurate data in the  $M_\pi$ and $F_\pi$ channels. 

When $M_\pi$ and $F_\pi$ are fitted {\em separately} in the range of 
 the four lowest fermion masses  closest to the critical surface, we get reasonable $\chi^2$ values for the fits,
 as shown in Figure~\ref{fig:sextetConformTest1}. However, the incompatibility of the fitted $\gamma$ values
disfavors the hypothesis,  inconsistent
with mass deformed conformal behavior.
\begin{figure}[h!]
\begin{center}
\begin{tabular}{c}
\includegraphics[height=5cm]{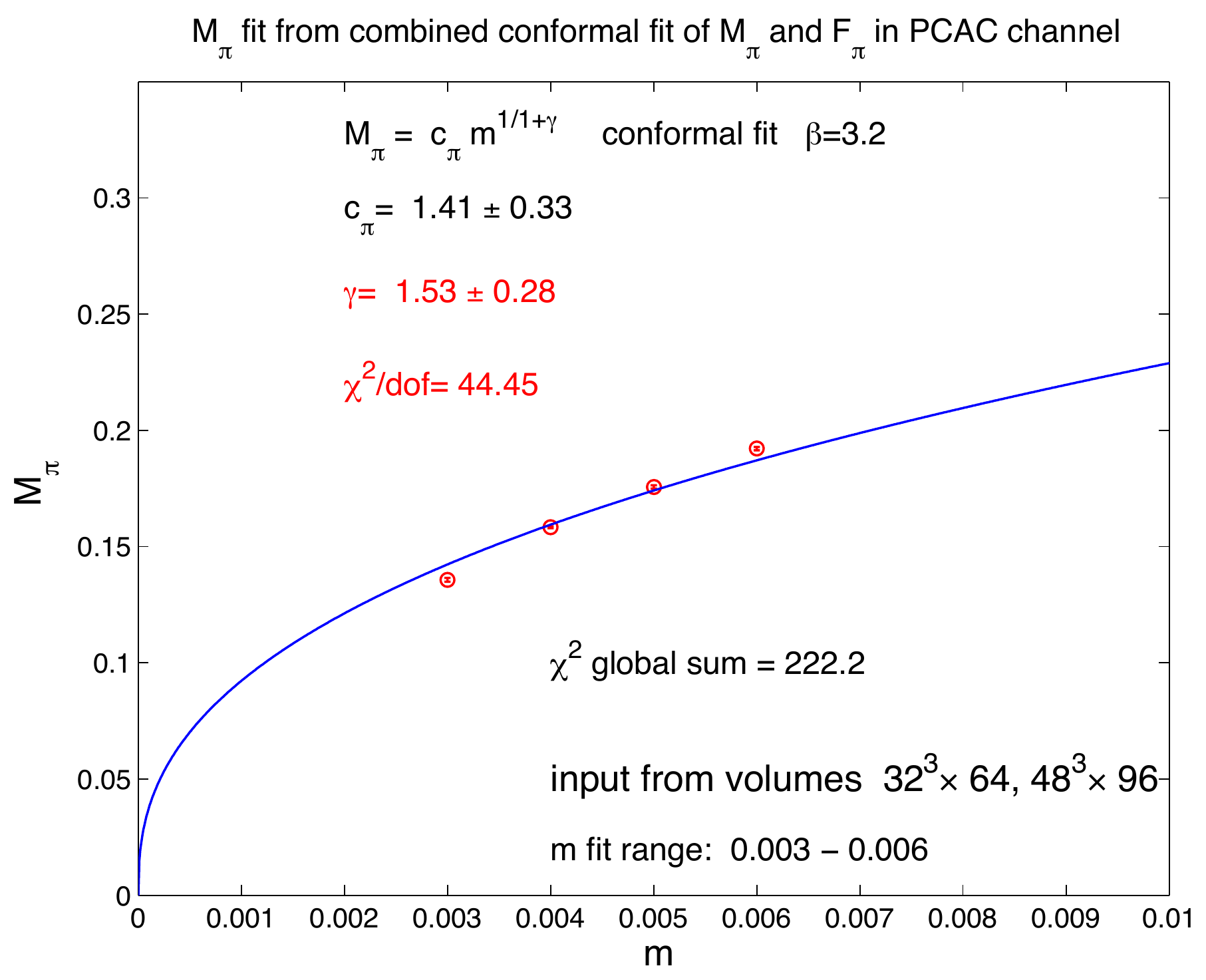}\\
\includegraphics[height=5cm]{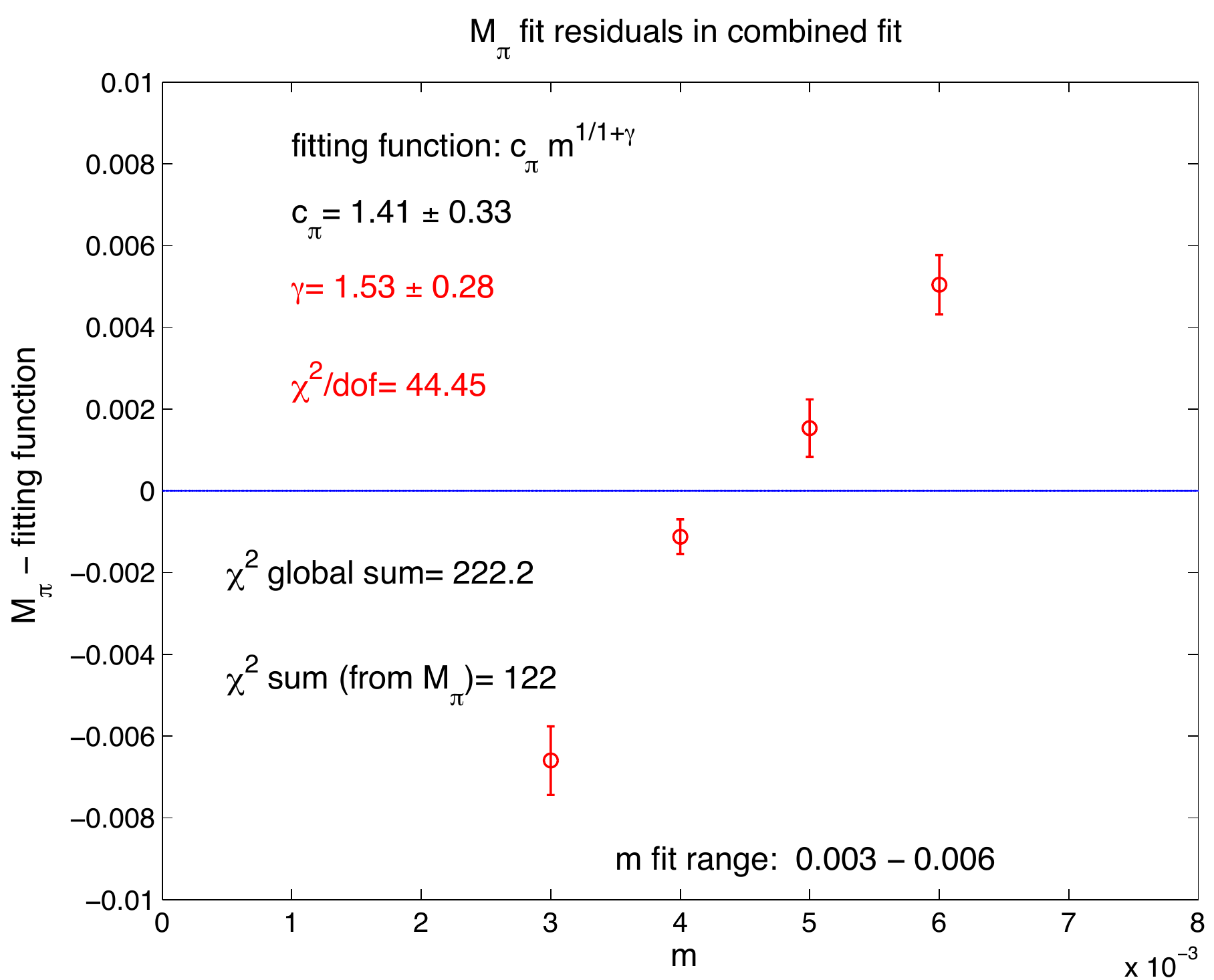}\\
\includegraphics[height=5cm]{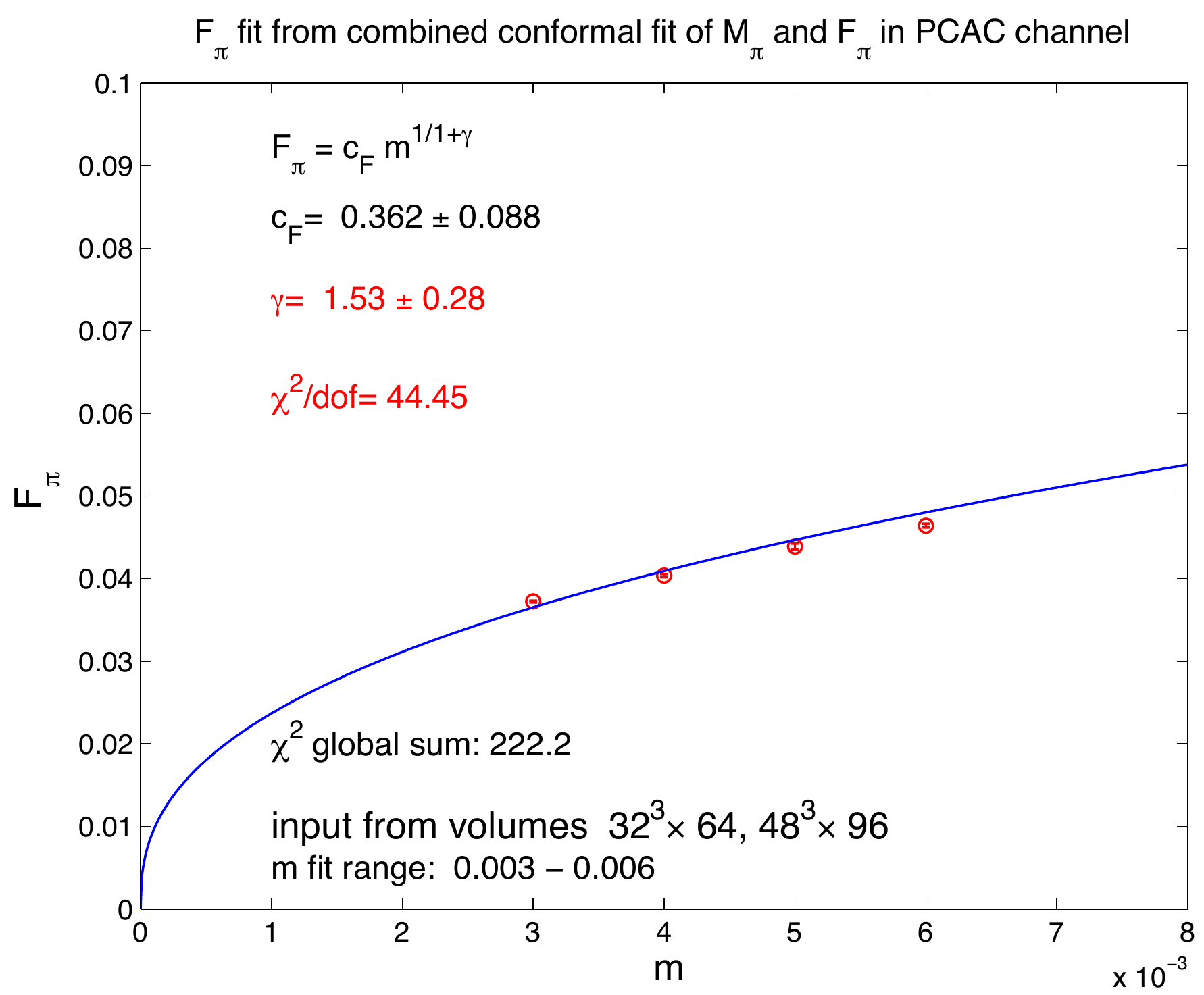}\\
\includegraphics[height=5cm]{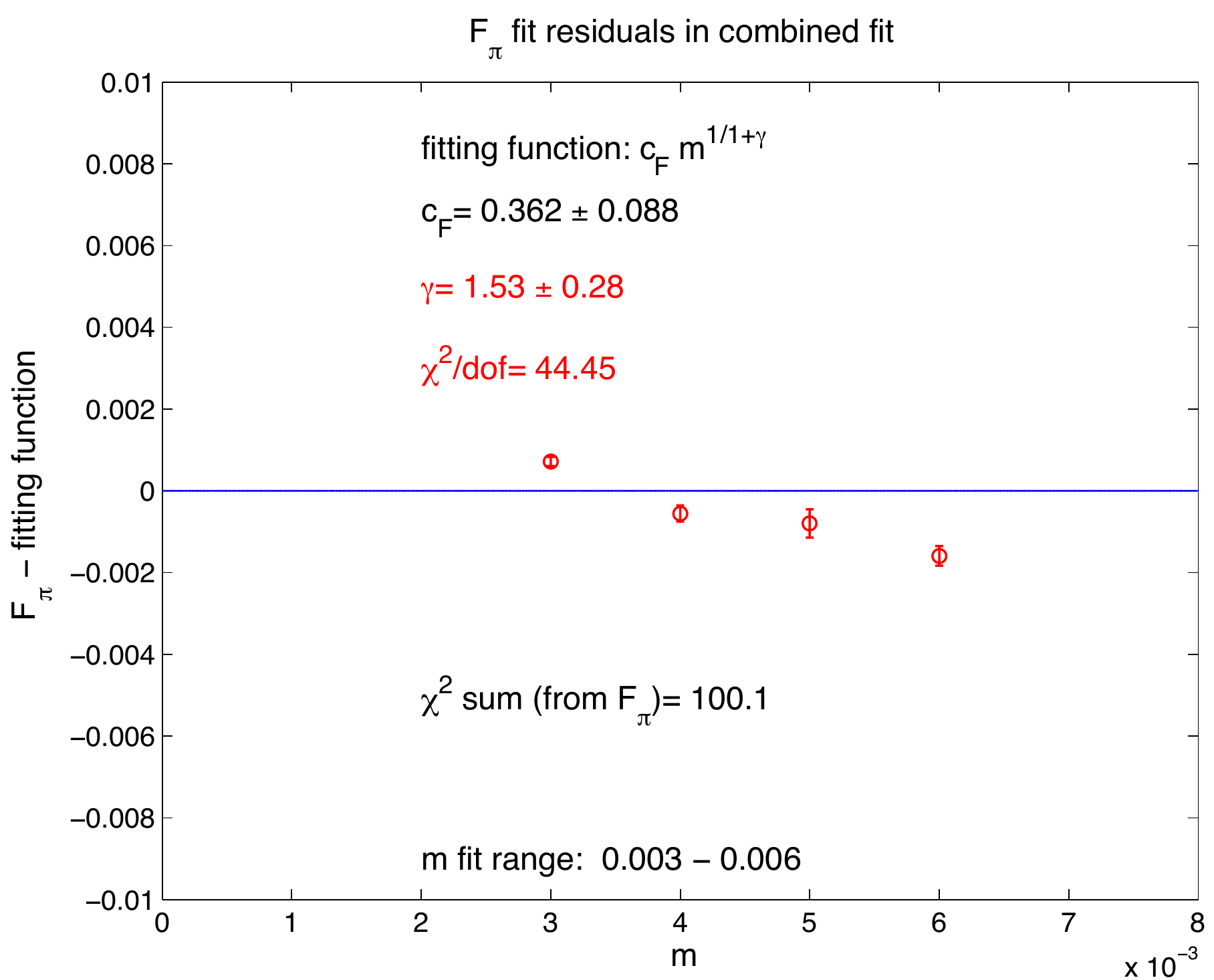}
\end{tabular}
\end{center}
\vskip -0.25in
\caption{\footnotesize The first plot shows the simultaneous conformal fit result for
the pion mass, while the second displays the $M_pi$ residuals. The last two
plots show the simultaneous fit result for the pion decay constant and the
$F_pi$ residuals."
The combined fit forces $\gamma=1.53(28)$
with an unacceptable ${\rm \chi^2/dof}$ of 44.5.}
\label{fig:sextetConformTest2}
\vskip -0.2in
\end{figure}
 The conflicting simultaneous fits to  universal conformal form with the same $\gamma$  
 for the Goldstone pion and the $F_\pi$ decay constant
 are illustrated  in Figure~\ref{fig:sextetConformTest2}. Fitting to the pion mass separately requires $\gamma=1.040(73)$ while the 
separate $F_\pi$ fit is forcing  $\gamma=2.20(15)$. In the combined fit they compromise with  $\gamma=1.53(28)$
and the unacceptable ${\rm \chi^2/dof}$ of 44.5.
It is important to note that the exponent $\gamma$  for the fit to $M_\pi$ only is what 
$\chi{\rm SB}$ would prefer. The separate conformal exponent $\gamma$  for $F_\pi$ is large to force 
to the origin the linear string of data which extrapolate to a finite constant in $\chi{\rm SB}$.
This creates conflict with the universal exponent $\gamma$  in the conformal analysis.   

From the tests we were able to perform, the sextet model is consistent with $\chi {\rm SB}$ and inconsistent with conformal symmetry. 
It will require further investigations to show that subleading effects cannot alter this conclusion. 
We will consider comprehensive conformal finite size scaling (FSS) tests which do not rely on 
infinite-volume extrapolation in the scaling fits.  
Conformal FSS was extensively applied to a different much discussed model with twelve fermion flavors in the fundamental representation
of the SU(3) color gauge group~\cite{Fodor:2012uu}. These kinds of tests are at a preliminary stage in the sextet project requiring new runs and 
systematic analysis. The FSS analysis of the existing dataset of this paper, when smaller volumes are included, 
disfavors the conformal hypothesis similarly to what we just presented in the infinite-volume limit. 
It is difficult to reconcile  $\chi{\rm SB}$ and large exponents in the fermion mass dependence with the low value of $\gamma$ 
defined by the chiral condensate 
using the Sch\"odinger functional for massless fermions~\cite{DeGrand:2012yq}.

\section{The new sextet Higgs project}
If $\chi {\rm SB}$ of the sextet model is  confirmed in the massless fermion limit, its potential relevance for the
realization of the composite Higgs mechanism is self-evident. 
The three Goldstone pions of the model have the perfect match 
for  providing the longitudinal components of the $W^{\pm}$ and $Z$ bosons.
The remaining most important issues are:  (1) to calculate the mass of the
$0^{++}$ state when the disconnected part of correlator I in Table 1 of~\cite{Ishizuka:1993mt} is included; (2) the determination of the 
non-perturbative gluon condensate on the lattice to clarify the dilaton connection if the Higgs particle turns out to be light;
(3) a more precise determination of the running coupling for which we will deploy our
new method based on the gradient flow of the gauge field in finite volume~\cite{Fodor:2012he}. 
We will outline in some details the first and second issues.

\subsection{The $ f_0$ state in the $0^{++}$ channel}

\begin{figure}[h!]
\begin{center}
\begin{tabular}{c}
\includegraphics[height=6cm]{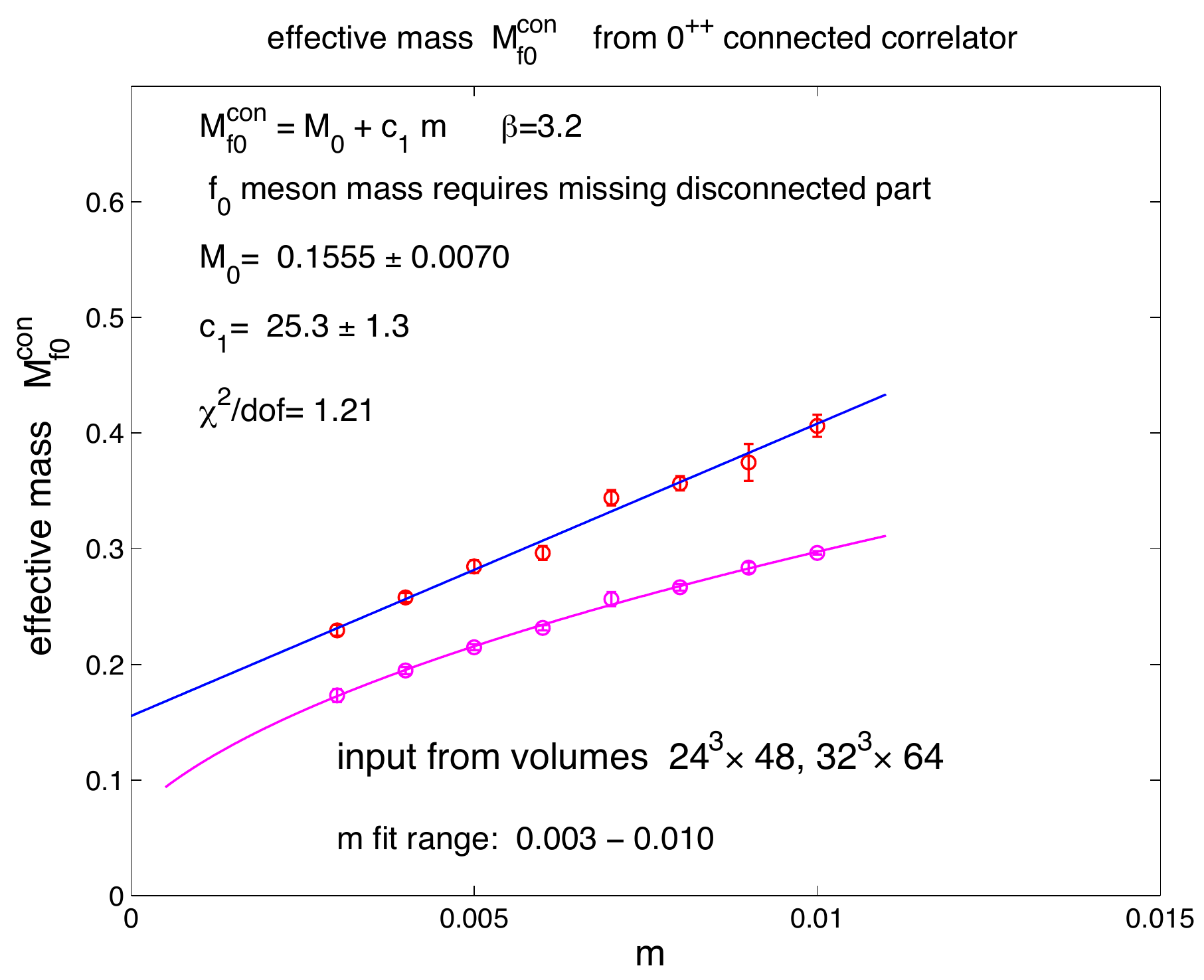}
\end{tabular}
\end{center}
\vskip -0.2in
\caption{\footnotesize  The linear fit is shown to the mass of the $0^{++}$ $f_0$ meson from the 
connected part of correlator I in Table 1 of~\cite{Ishizuka:1993mt}. For comparison, the scPion which is the parity partner of the $f_0$ meson 
in the correlator is replotted with its fit from Figure~\ref{fig:non-GoldstoneSpectrum} (magenta color). In the continuum limit,
the mass of the non-Goldstone scPion will vanish and the $f_0$ state could become light close to the conformal window.
The disconnected part of the correlator is required to resolve this issue.}
\vskip -0.1in
\label{fig:f0}
\end{figure}

Figure~\ref{fig:f0} shows the fermion mass dependence of the $f_0$ meson without including the disconnected
part of correlator I in Table 1 of~\cite{Ishizuka:1993mt}. The non-Goldstone scPion and $f_0$ are parity partner states in this correlator.
The quantum numbers of the $f_0$ meson match that of the $0^{++}$ state in the staggered correlator.
Close to the conformal window the $f_0$ meson is not expected to be similar to the $\sigma$ particle of QCD.
The full $f_0$ state including the disconnected diagrams could replace the role of the 
elementary Higgs and act as the Higgs impostor if it turns out to be light. 
It is  very difficult to do the full calculation including the disconnected diagram which is
the main part of our next generation sextet Higgs project. First, we will discuss preliminary  results  which ignore 
the disconnected part. The challenges will be outlined in the effort to include the disconnected part.

The linear  fit from the connected diagram is shown in Figure~\ref{fig:f0}. It has a non-zero intercept in the chiral limit with a mass more 
than five times $F$ so it corresponds to a heavy state and not a Higgs candidate. 
Since the $f_0$ state is the parity partner of the non-Goldstone scPion in the full correlator, the two states 
would become degenerate in the chiral limit with unbroken symmetry. Close to the conformal window it is reasonable to expect that the disconnected
diagram will dramatically reduce the $f_0$ mass and its split from the scPion when the chiral limit is taken. This
will leave the full $f_0$ state a viable Higgs candidate before new simulations resolve the issue and perhaps eliminate this attractive scenario.

To study flavor-singlet mesons, we need to consider fermion loops which are disconnected (often called hairpin diagrams).
Flavor-singlet correlators have 
fermion-line connected and fermion-line disconnected contributions from the hairpin diagrams.
To evaluate disconnected quark loops with zero momentum, we need to sum over propagators from sources 
at each spatial location for a given time slice. 
To avoid the very costly $\mathcal{O}(V)$
inversions to compute  all-to-all propagators in lattice terminology,  random sources have to be used with  noise
reduction.

A  very interesting further challenge and complication is the existence of two types of distinct $0^{++}$ scalar mesons. 
One of them is the composite fermion state and the other is the scalar glueball with the same quantum number.  
In dynamical sextet simulations, these two types of state will mix with an observable spectrum of scalar mesons which will 
require a well-chosen variational operator set to disentangle the scalar state. This further underlines the room left for a light
scalar state to emerge in the spectrum. It is also entirely possible that careful lattice calculations will shut down the Higgs interpretation.

Staggered fermions present an additional complication from the contribution of
pairs of pseudoscalar meson taste channels contributing to the scalar meson correlator.
To be a physical state, the scalar meson $f_0$ has to be taste singlet. Taste selection rules then require that the $f_0$ meson 
couples only to pairs of pseudoscalar mesons of the same taste. 
We have shown earlier in Section 4 that the pion taste multiplet splits into the Goldstone state and a variety of 
higher-lying non-Goldstone states, all degenerate 
with vanishing mass in the continuum limit. 
In the continuum limit only the taste singlet states (physical states) are expected to have the correct masses from the 
U(1) axial anomaly which is itself a taste singlet.  
The other non-singlet states remain light and create complicated threshold effects.
This complication is present in the $f_0$ correlator masked by the physical two-pion intermediate state~\cite{Bernard:2007qf}.

\subsection{The Higgs particle and the dilaton}
If the sextet model is  very close to the conformal window
with a small but nonvanishing $\beta$-function, a necessary 
condition is satisfied for spontaneous  breaking of scale invariance generating the light pseudo-Goldstone dilaton state. 
The model, as we argued earlier, is  also consistent with chiral symmetry breaking ($\chi{\rm SB}$) 
with the minimal Goldstone pion spectrum required for electroweak 
symmetry breaking and the Higgs mechanism. 
The very small beta function (walking) and $\chi{\rm SB}$ are
not sufficient to guarantee a light dilaton state if scale symmetry breaking and $\chi{\rm SB}$
are entangled in a complicated way. However, 
 a light Higgs-like scalar could emerge near the conformal window as a composite state, not necessarily with dilaton interpretation.
 To understand the important role of the non-perturbative gluon condensate in the partially 
 conserved dilatation current (PCDC) relation and its related dilaton implications, 
lattice simulations of the non-perturbative gluon condensate will be needed near the conformal window.

%
For discussion of the PCDC relation constraining 
the properties of the dilaton, we will closely follow the standard
argument like in~\cite{Appelquist:2010gy,Hashimoto:2010nw,Matsuzaki:2012fq}. We will also show how 
non-perturbative lattice methods
can explore the implications of the PCDC relation when applied to the sextet model. 

In strongly interacting gauge theories, like the sextet model under consideration, a dilatation current 
${\mathcal D}^\mu=\Theta^{\mu\nu}x_\nu$ 
can be defined from the symmetric energy-momentum tensor $\Theta^{\mu\nu}$. Although the massless theory is scale invariant on the classical
level, from the scale anomaly the dilatation current has a non-vanishing divergence,
\be
\partial_\mu \mathcal {D}^\mu = \Theta_\mu^\mu=\frac{\beta (\alpha)}{4\alpha}G^a_{\mu\nu}G^{a\mu\nu} \, .
\label{eq:D1}
\ee
Although $\alpha(\mu)$ and $G^a_{\mu\nu}G^{a\mu\nu} $ depend on the renormalization scale $\mu$, the trace of the 
energy-momentum tensor is scheme independent after renormalization. In the sextet model, the massless fermions are in the
two-index symmetric representation of the SU(3) color gauge group. The  gluon fields are
in the adjoint representation with  $G^a_{\mu\nu},~ a=1,2,...8$.
We will assume that the perturbative parts of the composite gauge operator $G^a_{\mu\nu}G^{a\mu\nu}$ and $\Theta_\mu^\mu$ 
are removed 
in Eq.~(\ref{eq:D1}) and only the non-perturbative (NP) infrared part will be considered in what follows. 

The dilaton coupling $f_\sigma$ is defined by the matrix element
\be
\langle 0| \Theta^{\mu\nu}(x)|\sigma(p)\rangle = \frac{f_\sigma}{3}(p^\mu p^\nu - g^{\mu\nu} p^2) e^{-ipx}
\label{eq:D2}
\ee
with $p^2=m^2_\sigma$ for the on-shell dilaton state $\sigma (p)$.
From the divergence of the dilatation current in Eq.~(\ref{eq:D1}) we get 
\be
  \langle 0| \partial_\mu\mathcal{D}^\mu(x)|\sigma(p)\rangle = f_\sigma m^2_\sigma  e^{-ipx}\, .
 \label{eq:D3}
\ee
The subtracted non-perturbative part of the energy-momentum tensor,
\be
\Bigl [\Theta^\mu_\mu \Bigr]_{NP} = \frac{\beta (\alpha)}{4\alpha}\Bigl[ G^a_{\mu\nu}G^{a\mu\nu}\Bigr]_{NP}\,,
 \label{eq:D3A}
\ee
 is defined by removing the perturbative part of the gluon condensate in the vacuum, 
\be
 \Bigl [\Theta^\mu_\mu \Bigr]_{NP} = \frac{\beta (\alpha)}{4\alpha}  G^a_{\mu\nu}G^{a\mu\nu} - 
 \langle 0|\frac{\beta (\alpha)}{4\alpha}G^a_{\mu\nu}G^{a\mu\nu}|0\rangle_{PT}\, .
 \label{eq:D4}
\ee 
The lattice implementation of the subtraction procedure will be briefly described after the derivation of the PCDC relation.
 
 It is easy to derive, like for example in~\cite{Appelquist:2010gy}, the dilaton matrix element 
of the energy-momentum tensor trace using some particular definition
 of the subtraction scheme,
 \be
 \langle \sigma (p=0)|\Bigl[\Theta^\mu_\mu (0)\Bigr]_{NP}|0\rangle \simeq \frac{4}{f_\sigma}\langle 0|\Bigl[\Theta^\mu_\mu (0)\Bigr]_{NP}|0\rangle \, .
 \label{eq:D5}
 \ee
When combined with Eq.~(\ref{eq:D3}), the partially conserved dilatation current (PCDC) relation is obtained,
\be
m^2_\sigma\simeq - \frac{4}{f^2_\sigma }\langle 0|\Bigl[\Theta^\mu_\mu (0)\Bigr]_{NP}|0\rangle \, .
\label{eq:D6}
\ee
Predictions for $m_\sigma$ close to the conformal window depend on the behavior 
of $f_\sigma$ and the gluon condensate $G^a_{\mu\nu}G^{a\mu\nu}$ of  Eq.~(\ref{eq:D3A}). There are two distinctly
different expectations about the limit of the gluon condensate to $f_\sigma$ ratio when the conformal window is approached.
In one interpretation,  the right-hand side is predicted to approach zero in the limit, so that the dilaton mass would parametrically vanish when
the conformal limit is reached ~\cite{Appelquist:2010gy}. 
The formal parameter is the non-physical (fractional) critical number of fermions  when the conformal phase is reached.
In an alternate interpretation the right-hand side ratio of Eq.~(\ref{eq:D6}) remains finite in the limit and a residual dilaton mass is expected~\cite{Hashimoto:2010nw,Matsuzaki:2012fq}. The two interpretations make different assumptions about the
entanglement of $\chi{\rm SB}$ and scale symmetry breaking but both scenarios expect a light dilaton mass 
in some exact non-perturbative realization of a viable BSM model.

It is important to note that there is no guarantee, even with a very small $\beta$-function near the conformal window, for
the realization of  a light enough dilaton to act as the new Higgs-like particle. Realistic BSM models have not been built with 
parametric tuning  close to the conformal window. For example, the sextet model is at some intrinsically determined position
near the conformal window and
only non-perturbative  lattice calculations can explore the physical properties of the scalar particle.

 \subsection{The non-perturbative gluon condensate on the lattice}


Power divergences are severe in the 
calculation of the lattice gluon condensate, 
because the operator $\alpha G^a_{\mu\nu}G^{a\mu\nu}$ has quartic divergences.
The gluon condensate is computed on the lattice from the 
expectation value of the plaquette operator $U_P$. On the tree level we have the relation 
\begin{equation}
{\rm lim}_{a\rightarrow 0} ~~\Biggl (\frac{1}{a^4}\, \langle 1-\frac{1}{3}\,\mbox{tr}\,U_P\rangle \Biggr )
= \frac{\pi^2}{36} \,\langle\frac{\alpha}{\pi} 
GG\rangle_{\rm lattice} 
\label{eq:plaquette1}
\end{equation}
as the continuum limit is approached in the limit of vanishing bare lattice coupling $g_0$.
At finite lattice coupling we have the sum of a perturbative series in  $g_0$ and the non-perturbative gluon condensate,
\begin{equation} 
\Bigl\langle 1-\frac{1}{3}\,\mbox{tr}\,U_P\Bigr\rangle
= \sum_{n}c_n \cdot g_0^{2n} +
a^4\,\frac{\pi^2}{36} \,\Biggl( \frac{b_0}{\beta(g_0)}  \Biggr)\,\Bigl\langle\frac{\alpha}{\pi} 
GG\Bigr\rangle_{\rm lattice} + \,O(a^6)\, ,
\label{eq:plaquette2}
\end{equation}
where $b_0$ is the leading $\beta$-function coefficient.
There is no gauge-invariant operator of dimension 
2 and therefore the order $a^2$ term is missing in Eq.~(\ref{eq:plaquette2}). 
For small lattice spacing $a$, the perturbative
series is much larger than the non-perturbative gluon condensate, and its determination requires the subtraction of
the perturbative series from the high accuracy Monte Carlo data of the plaquette. 
The $c_n$ expansion coefficents can be determined to high order using stochastic perturbation theory~\cite{Di Renzo:2004ge}.
This procedure  requires  the investigation of Borel summation of the high order terms in the perturbative expansion since the coefficients $c_n$
are expected to diverge in factorial order and one has to deal with the well-known renormalon issues. 
The methodology has been extensively studied in pure Yang-Mills theory
on the lattice~\cite{Horsley:2012ra}. 

It will be very important to undertake similar investigations of the non-perturbative gluon
condensate in the sextet model with full fermion dynamics.
We hope to return to this problem in the near future.


\section*{Summary and outlook}
We have shown that the chiral condensate and the mass spectrum of the sextet model 
are consistent with chiral symmetry breaking in the limit of vanishing fermion mass. 
In contrast, 
sextet fermion mass deformations of 
spectral properties are not consistent with leading conformal scaling behavior near the critical surface 
of a conformal theory.
Our new results are reconciled with recent findings of the sextet $\beta$-function~\cite{DeGrand:2012yq},
if the model is close to the conformal window with a  very small non-vanishing $\beta$-function. 
This leaves open the possibility
of a light scalar state with quantum numbers of the Higgs impostor. The light Higgs-like state could emerge as the 
pseudo-Goldstone dilaton from spontaneous symmetry breaking of scale invariance. 
Even without association with the dilaton, the scalar Higgs-like state 
can be light if the sextet gauge model is very close to the conformal window.
A new Higgs project of sextet lattice simulations was outlined to resolve these important questions.
Plans include the determination of the S parameter and the sextet confining force
with results on the string tension already
reported, strongly favoring the $\chi{\rm SB}$ hypothesis~\cite{Holland}. 

\section*{Acknowledgments}
This work was supported by the DOE under grant DE-FG02-90ER40546,
by the NSF under grants 0704171 and 0970137, by the EU Framework Programme 7 grant (FP7/2007-2013)/ERC
No 208740, and by the Deutsche Forschungsgemeinschaft grant SFB-TR 55.
The simulations were performed using USQCD computational resources
at Fermilab and JLab.  Further support was provided by the UCSD GPU cluster 
funded by DOE ARRA Award  ER40546.
Some of the simulations used allocations from 
the Extreme Science and Engineering Discovery Environment (XSEDE), 
which is supported by National Science Foundation grant number OCI-1053575.
In addition, some computational resources were used at the University of Wuppertal, Germany. 
We are grateful to Kalman Szabo and Sandor Katz
for their code development building on Wuppertal gpu technology~\cite{Egri:2006zm}. 
KH wishes to thank the Institute for Theoretical Physics and the 
Albert Einstein Center for Fundamental Physics at Bern University for their support.


\end{document}